\documentclass[preprint,12pt]{elsarticle}

\usepackage{hyperref}
\usepackage{booktabs}
\usepackage{graphicx}
\usepackage{subfigure}
\usepackage{float}
\usepackage[export]{adjustbox}
\usepackage{dcolumn}
\usepackage{amsmath}
\usepackage{amssymb}
\usepackage{bm}
\usepackage{cleveref}
\usepackage[utf8]{inputenc}
\usepackage[T1]{fontenc}
\usepackage{threeparttable}
\usepackage{etoolbox}
\usepackage{color}
\usepackage{xcolor}
\usepackage{comment}
\usepackage{multirow}
\usepackage{tikz}
\usetikzlibrary{arrows.meta,positioning,calc,decorations.pathreplacing}
\usepackage{lineno}
\modulolinenumbers[5]

\journal{International Journal of Heat and Mass Transfer}

\begin{document}

\begin{frontmatter}

\title{Conservative local grid refinement for the vectorial lattice Boltzmann method}

\author[eth]{S.A.~Hosseini\corref{cor1}}
\ead{shosseini@ethz.ch}
\author[eth]{R.M.~Str\"assle}
\ead{rubenst@ethz.ch}
\author[eth]{I.V.~Karlin}
\ead{ikarlin@ethz.ch}

\cortext[cor1]{Corresponding author}
\address[eth]{Computational Kinetics Group, Department of Mechanical and Process Engineering, ETH Z\"urich, 8092 Zurich, Switzerland}

\begin{abstract}
We present a conservative local grid-refinement coupling for the vectorial lattice Boltzmann method (VLBM), a kinetic-type scheme for hyperbolic conservation laws in which each conserved field is carried by its own population rather than by moment truncation of a Maxwellian. Coarse and fine regions share a common link speed and are advanced with different time steps, sub-cycled at a fixed ratio; a mean-preserving reconstruction in time supplies the missing coarse-to-fine boundary data, and a matched, Berger--Colella-style reflux returns the fine-to-coarse data, together conserving mass, momentum, and energy exactly, independent of the equation of state, the local relaxation parameter, or the time step. In addition, we propose an adaptive mesh refinement extension through a conservative pair of splitting and merging operators that require no modification to the underlying coupling. The approach is validated on a sequence of increasingly demanding two-dimensional cases, in every case recovering exact conservation and the expected accuracy gains over an equivalent uniform grid.
\end{abstract}


\end{frontmatter}


\section{Introduction}
\label{sec:Introduction}
Compressible flows -- transonic and supersonic aerodynamics, blast and shock waves, high-speed reacting and astrophysical flows -- are intrinsically multi-scale: a shock front, a contact discontinuity, or a thin shear layer can occupy a vanishing fraction of the domain while controlling the accuracy of the solution everywhere else; Resolving the entire domain at the resolution required by those features, even in surrounding, smoothly varying regions would be prohibitively expensive. Finite-volume and finite-difference schemes built around approximate or exact Riemann solvers have long been the workhorse for this class of problem~\cite{Godunov1959,Roe1981,vanLeer1979,LeVeque2002,Toro2009}, and local grid refinement -- resolving only the small region that actually needs it, at higher resolution, while leaving the bulk of the domain coarse -- has been one of the oldest and most effective ways to recover that accuracy without paying for a globally fine mesh~\cite{BergerOliger1984,BergerColella1989}.

The lattice Boltzmann method (LBM) is a solver that has emerged as an attractive alternative to these classical solvers, owing to its low per-cell computational cost, the purely local and on-grid streaming-collision structure of its update -- which makes it parallel and exactly conservative -- and a kinetic origin that recovers the target macroscopic equations with minimal numerical dissipation~\cite{ChenDoolen1998,Succi2001}. Its success in the incompressible regime is well established; extending it to compressible flows has been an active research direction, producing a range of formulations -- from pure LBM schemes~\cite{Frapolli2015,saadat2021extended,HosseiniFeinbergKarlin2026} to hybrid LBM/finite-volume or LBM/finite-difference couplings~\cite{feng2019hybrid,guo2020efficient} -- that have been shown to successfully model compressible flows across the subsonic, transonic, and mildly supersonic regimes~\cite{Bayat2026}.

A numerical model formally of the same family, but vectorial rather than scalar in its populations, has recently re-emerged as an alternative for hyperbolic systems of partial differential equations~\cite{JinXin1995,Bouchut1999,Bouchut2004,Graille2014,Dubois2014},
attracting renewed interest in recent years~\cite{GuillonHelieHelluy2024,WissocqLiuAbgrall2025,AregbaDriolletBellotti2025,bukreev2026lattice}. The vectorial lattice Boltzmann method (VLBM) keeps the properties that make LBM attractive in the first place -- an on-lattice, exact streaming operator, low computational cost etc-- but departs from it structurally: rather than every population being a scalar, truncated-Maxwellian moment of the full conserved state, each conservation law of the target hyperbolic system is carried by its own population, with the vector equilibrium fixed by exact algebraic consistency with the target flux rather than by moment truncation of the Maxwell-Boltzmann distribution. Extending the scheme to the more practical settings a real simulation needs, however, remains largely unexplored. The one exception the authors are aware of is a multiresolution, wavelet-based adaptive framework developed for the same family of relaxation-type schemes~\cite{Bellotti2022JCP,Bellotti2022CRAS}, which dynamically coarsens and refines the mesh under an a priori error control. In a recent publication, we proposed a consistent approach to adaptive time-stepping for VLBM~\cite{VLBMCompressiblePaper}; here we propose, further, a simple model for local grid refinement that is both consistent, and exactly conservative under the same vectorial construction, along with an extension to adaptive mesh refinement.

The present paper is organized as follows: \S\ref{sec:vlbm} recalls the VLBM scheme used as bulk solver; \S\ref{sec:refinement} develops the refinement coupling itself, first in one dimension (\S\ref{sec:1d}) -- introducing the coarse-to-fine reconstruction and fine-to-coarse reflux operations-- then in two dimensions (\S\ref{sec:2d}), extending the construction to 2-D refinement patches, and finally to adaptive mesh refinement (\S\ref{sec:amr}), through a conservative pair of splitting and merging operations; \S\ref{sec:validation} presents a variety of numerical simulations probing the validity and consistency of the proposed refinement strategy, closing with a fully dynamic, multi-level adaptive run.

\section{Vectorial lattice Boltzmann}
\label{sec:vlbm}
This section recalls, without derivation, the elements of the VLBM used here as bulk solver; Details of the solver can be found in~\cite{VLBMCompressiblePaper}.
\subsection{Target balance equations}
\label{sec:vlbm-balance}
The target is the compressible Euler system for a single-component, inviscid fluid with conserved vector $\bm W=(\rho,\rho\bm u,\rho E)^\dagger\in\mathbb R^{D+2}$ ($D$ the spatial dimension, $\rho$ the density, $\bm u$ the velocity, and $E$ the specific total energy), written compactly as
\begin{equation}
  \partial_t \bm W + \nabla\cdot \bm Q(\bm W) = \bm 0,
  \label{eq:bal-general-form}
\end{equation}
with $\bm Q(\bm W)$ the flux. The pressure entering $\bm Q$ is related to the thermodynamic state through an equation of state left \emph{generic} throughout --in numerical application only ideal gas cases are presented, subject only to the hyperbolicity requirement that the system's characteristic speeds be real; this genericity is what allows the same scheme, and the same grid-refinement coupling, to be applied unmodified to an ideal gas or a van der Waals fluid alike, as demonstrated for the base scheme in~\cite{VLBMCompressiblePaper} and inherited without change by the refinement coupling of \S\ref{sec:refinement}, since that coupling never references the equation of state explicitly.

\subsection{Vectorial LBM}
\label{sec:vlbm-model}
Unlike scalar-population lattice Boltzmann schemes, whose equilibria are truncated-Maxwellians, the VLBM carries a share of the \emph{full} conserved vector $\bm W$ directly in each population $\bm f_i(\bm x,t)\in\mathbb R^{D+2}$, with the equilibrium fixed by exact algebraic consistency with the target flux rather than by moment truncation~\cite{JinXin1995,Bouchut1999,Bouchut2004,Graille2014,Dubois2014}. We restrict throughout to the $DdQ2^d$ family: $Q=2^D$ discrete velocities, arranged as $D$ antipodal pairs,
\begin{equation}
    \bm c_{\alpha,\pm}=\pm c\,\bm e_\alpha (\alpha=1,\dots,D),
\end{equation}
aligned with the Cartesian directions, with a single, \emph{isotropic} link speed $c=\delta x/\delta t$ common to every direction and every population.
Fixing $c$ this way makes the transport sub-step an exact grid shift, free of numerical dissipation or dispersion by construction. The consistency conditions,
\begin{equation}
    \sum_i\bm f_i^{\rm eq}=\bm W,\, \sum_i \bm c_i\bm
f_i^{\rm eq}=\bm Q(\bm W)
\end{equation}
fix the equilibrium exactly,
\begin{equation}
  \bm f_i^{\rm eq}(\bm W) = \frac{\bm W}{2D} + \frac{\bm c_i\cdot\bm
  Q(\bm W)}{2c^2}.
  \label{eq:num-equilibrium}
\end{equation}
One time step is the exact transport sub-step,
\begin{equation}
    \bm f_i(\bm x,t)=\bm f_i(\bm x-\delta t\,\bm c_i,\,t-\delta t),
\end{equation}
followed by a Bhatnagar-Gross-Krook (BGK) relaxation sub-step,
\begin{equation}
  \bm f_i(\bm x,t) = 2\beta\,\bm f_i^{\rm eq}\big(\bm W(\bm x,t)\big) +
  (1-2\beta)\,\bm f_i(\bm x,t),
  \label{eq:num-relaxation}
\end{equation}
with $\bm W=\sum_i\bm f_i$ and relaxation parameter $\beta\in(0,1]$. Because $\sum_i\bm f_i^{\rm eq}(\bm W)=\bm W$ identically, the zeroth moment of \eqref{eq:num-relaxation} is invariant under relaxation for \emph{any} $\beta$, cell by cell: mass, momentum, and energy are conserved exactly at the discrete level regardless of $\beta$ or the equation of state. It is worth nothing that, as discussed in~\cite{VLBMCompressiblePaper} we will rely on acoustic scaling when refining the grid to ensure convergence to the Euler limit, i.e. $\delta x/\delta t=c={\rm cst}$. Further since a constant $\beta$ makes the scheme's leading dissipative correction to \eqref{eq:bal-general-form} formally of the same order as a genuine Navier--Stokes viscous stress, we use a space/time-adaptive $\beta$ driven by a sensor, introduced in~\cite{VLBMCompressiblePaper}. $\beta\to1$ (vanishing dissipation) wherever the flow is resolved and smooth, and towards $\beta\to\beta_{\min}$ only where a genuine, under-resolved compression is detected: for each conserved component $k$, a grid-consistent, dimensionless jump estimate 
\begin{equation}
    \hat s_k=\delta x\,(\nabla\cdot\bm Q)_k/\max_{\bm x,\alpha}|Q_{\alpha,k}|,
\end{equation}
is combined into a scalar activity
\begin{equation}
    \bar s=\mathcal S[(\tfrac1{D+2}\sum_k\hat s_k^2)^{1/2}],
\end{equation}
where $\mathcal S$ denotes a box-filtering operation to prevent Gibbs-like oscillations. The scalar activity is gated by a Ducros-type dilatation/vorticity discriminator $\bar\theta\in[0,1]$ built from,
\begin{equation}
    \mathrm{dil}=\nabla\cdot\bm u,
\end{equation}
and
\begin{equation}
    \mathrm{vort}=\partial_x u_y-\partial_yu_x,
\end{equation}
and (in 2D)so that smooth rotational structures are not mistaken for shocks. The final form of the space/time-adaptive $\beta$ is,
\begin{equation}
  \beta(\bm x) = \beta_{\max} - (\beta_{\max}-\beta_{\min})\,
  \mathrm{clip}\big(C_{\rm sensor}\,\bar s(\bm x)\,\bar\theta(\bm x),\,0,\,1\big),
  \label{eq:sensor-B}
\end{equation}
where $\mathrm{clip}(,0,1)$ is a standard function clipping the values of its argument to the $[0\,1]$ interval.
\section{Grid refinement}
\label{sec:refinement}
\subsection{Preliminary: streaming as a finite-volume flux update}
\label{sec:fv-basis}
Everything below -- the reconstruction of a missing fine-grid value in \S\ref{sec:1d}, its 2-D counterpart in \S\ref{sec:2d}, and the exact conservation-- rests on a single identity, established here once for an ordinary, single-resolution $D1Q2$ grid before any coarse--fine interface is introduced: streaming \emph{is} a special finite-volume flux update. Consider a uniform grid of spacing $\delta x$, cell index $j$, and write $f_+$ for the rightward $D1Q2$ population (consistent with $\bm f_i$ of \S\ref{sec:vlbm-model} restricted to $D=1$; the leftward population $f_-$ is identical by mirror symmetry). Between collisions, $f_+$ obeys the pure advection equation $\partial_t f_++c\,\partial_x f_+=0$; integrating this conservation law over a cell of width $\delta x$ and one sub-step $\delta t$, exactly as for any finite-volume discretization of a scalar advection law, gives
\begin{equation}
  f_{+,j}^{n+1} = f_{+,j}^{*,n} -
  \frac{\delta t}{\delta x}\Big(F_{+,j+1/2}-F_{+,j-1/2}\Big),
  \label{eq:1d-fv}
\end{equation}
with $f_{+,j}^{*,n}$ representing the post-collision discrete distribution function and
$F_{+,j-1/2}:=c\,f_+^*(x_{j-1/2}^-\,)$ the numerical flux through the cell's left face, evaluated from the upwind side by whatever reconstruction is chosen inside cell $j-1$. One possible reconstruction for the population at the interface is $f_+(x)\equiv f_{+,j-1}^{*,n}$ throughout cell $j-1$ over the sub-step, giving $F_{+,j-1/2}=c\,f_{+,j-1}^{*,n}$ whenever the face's domain of dependence $[x_{j-1/2}-c\,\delta t,\,x_{j-1/2}]$ stays inside cell $j-1$. Substituting into \eqref{eq:1d-fv},
\begin{equation}
  f_{+,j}^{n+1} = f_{+,j}^{*,n} -
  \frac{c\,\delta t}{\delta x}\Big(f_{+,j}^{*,n}-f_{+,j-1}^{*,n}\Big).
  \label{eq:1d-fv-nu}
\end{equation}
In the case of a lattice Boltzmann streaming, $c=\delta x/\delta t$, i.e. CFL=1, collapsing Eq.~\eqref{eq:1d-fv-nu} to
\begin{equation}
  f_{+,j}^{n+1}=f_{+,j-1}^{*,n},
\end{equation}
i.e.\ exact population-for-population shift. That equivalence is what the coarse--fine coupling below turns on. Wherever one side of an interface has not yet advanced to a given sub-step, what its neighbor is missing to complete \eqref{eq:1d-fv} is not a streamed value but a \emph{flux sample} -- reconstructing it is therefore a finite-volume closure problem, to be solved by the same logic (consistency plus conservation) that closes any finite-volume scheme. \S\ref{sec:1d} develops that closure in 1-D; \S\ref{sec:2d} reuses \eqref{eq:1d-fv} unchanged, one cartesian direction at a time, in 2-D.

\subsection{1-D refinement sequence}
\label{sec:1d}
\begin{figure}[t]
\centering
\definecolor{cL0}{RGB}{237,231,246}
\definecolor{cR0}{RGB}{255,224,213}
\definecolor{cRecon}{RGB}{0,121,107}
\definecolor{cReflux}{RGB}{21,101,192}
\begin{tikzpicture}[scale=0.85,every node/.style={font=\footnotesize}]
 \def\xL{1.0}
 \def\xR{4.4}
 \def\bw{1.9}
 \def\bh{1.1}
 \def\yA{7.2}
 \def\yB{3.9}
 \def\yC{0.6}
 \draw[-{Latex[length=1.8mm]}] (-0.3,8.9) -- (-0.3,-0.2);
 \node[anchor=east] at (-0.45,\yA) {$t_n$};
 \node[anchor=east] at (-0.45,\yB) {$t_n+\delta t_2$};
 \node[anchor=east] at (-0.45,\yC) {$t_n+\delta t_1$};
 \node[rotate=90,anchor=center] at (-1.6,3.9) {time};
 \node[font=\footnotesize\bfseries] at (\xL,8.7) {$L_0$ (coarse)};
 \node[font=\footnotesize\bfseries] at (\xR,8.7) {$R_0$ (fine)};
 \draw[thick,dashed,gray] (2.7,9.0) -- (2.7,-0.5);
 \node[above,gray] at (2.7,9.05) {interface};
 \draw[fill=cL0,thick] (\xL-\bw/2,\yA-\bh/2) rectangle (\xL+\bw/2,\yA+\bh/2);
 \draw[-{Latex[length=1.6mm]},thick] (\xL-0.75,\yA+0.25) -- (\xL+0.75,\yA+0.25);
 \node[above] at (\xL,\yA+0.85) {\scriptsize $f_{L_0,+}^{*}\equiv E_1$};
 \draw[-{Latex[length=1.6mm]},thick] (\xL+0.75,\yA-0.25) -- (\xL-0.75,\yA-0.25);
 \node[below] at (\xL,\yA-0.68) {\scriptsize $f_{L_0,-}(t_n)$};
 \draw[fill=cR0,thick] (\xR-\bw/2,\yA-\bh/2) rectangle (\xR+\bw/2,\yA+\bh/2);
 \draw[-{Latex[length=1.6mm]},thick] (\xR-0.75,\yA+0.25) -- (\xR+0.75,\yA+0.25);
 \node[above] at (\xR,\yA+0.85) {\scriptsize $f_{R_0,+}(t_n)$};
 \draw[-{Latex[length=1.6mm]},thick] (\xR+0.75,\yA-0.25) -- (\xR-0.75,\yA-0.25);
 \node[below] at (\xR,\yA-0.68) {\scriptsize $f_{R_0,-}(t_n)$};
 \draw[fill=cR0,thick] (\xR-\bw/2,\yB-\bh/2) rectangle (\xR+\bw/2,\yB+\bh/2);
 \draw[cRecon,-{Latex[length=1.6mm]},thick] (\xR-0.75,\yB+0.25) -- (\xR+0.75,\yB+0.25);
 \node[above,cRecon] at (\xR,\yB+0.68) {\scriptsize $f^{\rm recon}_{R_0,+}(k{=}1)$};
 \draw[-{Latex[length=1.6mm]},thick] (\xR+0.75,\yB-0.25) -- (\xR-0.75,\yB-0.25);
 \node[below] at (\xR,\yB-0.68) {\scriptsize $f_{R_0,-}^{*,1}$};
 \draw[fill=cL0,thick] (\xL-\bw/2,\yC-\bh/2) rectangle (\xL+\bw/2,\yC+\bh/2);
 \draw[-{Latex[length=1.6mm]},thick] (\xL-0.75,\yC+0.25) -- (\xL+0.75,\yC+0.25);
 \node[above] at (\xL,\yC+0.68) {\scriptsize $f_{L_0,+}^{*}(t_n{+}\delta t_1)$};
 \draw[cReflux,-{Latex[length=1.6mm]},thick] (\xL+0.75,\yC-0.25) -- (\xL-0.75,\yC-0.25);
 \node[below,cReflux] at (\xL,\yC-0.68) {\scriptsize $f_{L_0,-}^{\rm new}$};
 \draw[fill=cR0,thick] (\xR-\bw/2,\yC-\bh/2) rectangle (\xR+\bw/2,\yC+\bh/2);
 \draw[cRecon,-{Latex[length=1.6mm]},thick] (\xR-0.75,\yC+0.25) -- (\xR+0.75,\yC+0.25);
 \node[above,cRecon] at (\xR,\yC+0.68) {\scriptsize $f^{\rm recon}_{R_0,+}(k{=}2)$};
 \draw[-{Latex[length=1.6mm]},thick] (\xR+0.75,\yC-0.25) -- (\xR-0.75,\yC-0.25);
 \node[below] at (\xR,\yC-0.68) {\scriptsize $f_{R_0,-}^{*,2}$};
 \draw[cRecon,dashed,thick,-{Latex[length=1.6mm]}]
   (\xL+0.75,\yA+0.25) .. controls (3.2,6.0) and (3.2,4.4) .. (\xR-0.75,\yB+0.25);
 \draw[cRecon,dashed,thick,-{Latex[length=1.6mm]}]
   (\xL+0.75,\yA+0.25) .. controls (3.5,5.0) and (3.5,1.0) .. (\xR-0.75,\yC+0.25);
 \draw[cReflux,thick]
   (\xR-0.75,\yB-0.25) .. controls (2.0,2.5) and (1.6,1.0) .. (\xL+0.75,\yC-0.25);
 \draw[cReflux,thick]
   (\xR-0.75,\yC-0.25) .. controls (2.0,0.6) and (1.7,0.6) .. (\xL+0.75,\yC-0.25);
 \draw[gray] (5.5,6.5) rectangle (9.5,7.9);
 \node[font=\footnotesize\bfseries] at (7.5,7.65) {Legend};
 \draw[thick] (5.75,7.25) -- (6.2,7.25);
 \node[anchor=west] at (6.3,7.25) {known / ordinary};
 \draw[cRecon,dashed,thick] (5.75,6.9) -- (6.2,6.9);
 \node[anchor=west,cRecon] at (6.3,6.9) {reconstruction (Step 3)};
 \draw[cReflux,thick] (5.75,6.6) -- (6.2,6.6);
 \node[anchor=west,cReflux] at (6.3,6.6) {reflux (Step 4)};
\end{tikzpicture}
\caption{Space--time schematic of one coarse cycle at the interface between coarse cell $L_0$ and fine cell $R_0$, following the five steps of \S\ref{sec:1d}. Rows are time levels, columns are cells; $L_0$ only updates at $t$ and $t_n+\delta t_1$ (Step~2), while $R_0$
sub-cycles through $t_n+\delta t_2$ ($k=1$) to $t_n+\delta t_1$ ($k=2$).}
\label{fig:1d-schematic}
\end{figure}
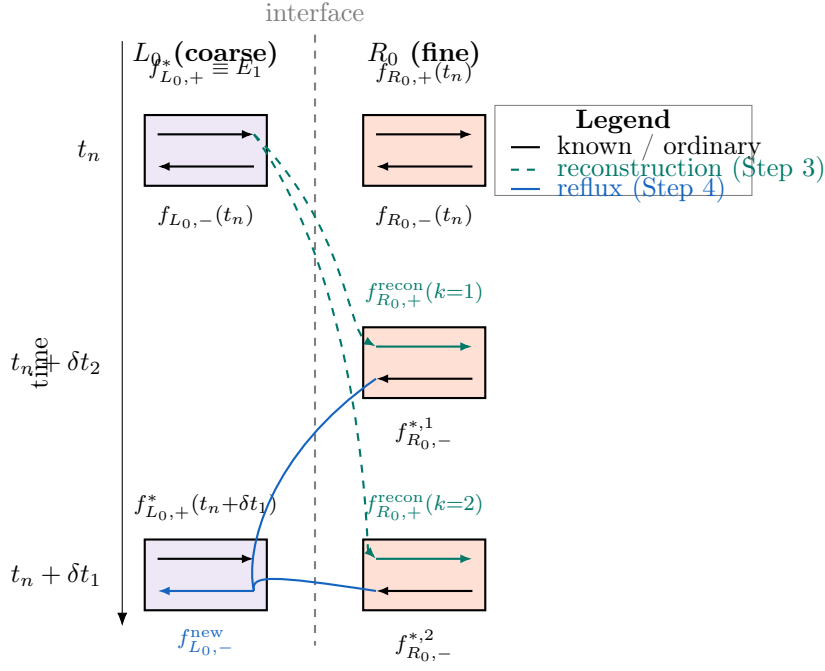

The construction proceeds in five steps, all summarized together in Fig.~\ref{fig:1d-schematic}. Note that all refinement discussions present in the context -- to converge to the target Euler limit -- of the present work assume acoustic scaling, i.e.\ constant $c$ regardless of the refinement level. The algorithm proceeds as follows: (a) collision on all grid-points, both coarse and fine, followed by streaming. This takes the fine domain to $t+\delta t_2$ and the coarse side to $t+\delta t_1$, leaving $f_+$ at the right-most coarse point denoted $L_0$ at $t+\delta t_1$ and $f_-$ at the left-most fine point denoted $R_0$ at $t+\delta t_2$ as unknowns. (b) Computation of missing population at $R_0$ and $t+\delta t_2$. (c) Conduct additional collision/streaming/reconstruction steps on the fine side taking it to $t+\delta t_1$. (d) As in step (b) fill missing population $f_+$ at $R_0$ at $t+\delta t_2$. (e) Finally, reconstruct missing population $f_-$ at $L_0$ and $t+\delta t_1$.
\subsubsection{Step 1: two resolutions, one link speed}
A coarse region of spacing $\delta x_1$ meets a fine region of spacing $\delta x_2=\delta x_1/r$ ($r\in\mathbb N$, $r\ge1$ the refinement ratio) at a single interface, coarse cell $L_0$ facing fine cells $R_0,R_1,\dots$ (Fig.~\ref{fig:1d-schematic}). We keep the link speeds identical --per eh acoustic scaling requirement, $c_1=c_2=c$, and let the time step differ instead,
\begin{equation}
  \delta t_1 = \frac{\delta x_1}{c}, \qquad
  \delta t_2 = \frac{\delta x_2}{c} = \frac{\delta t_1}{r}.
  \label{eq:1d-dtratio}
\end{equation}
Both sides then obey \emph{exactly} the same equilibrium~\eqref{eq:num-equilibrium} and
relaxation~\eqref{eq:num-relaxation}; they differ only in how often they are advanced, one coarse step against $r$ fine steps.
\subsubsection{Step 2: applying the flux identity at the interface}
\label{sec:1d-subcycle}
Write $f_{L_0,+}$, $f_{L_0,-}$ for the two $D1Q2$ populations of $L_0$ and $f_{R_0,+}$, $f_{R_0,-}$ for those of $R_0$ immediately across the interface (subscripts $+,-$ for the rightward/leftward link, consistent with $\bm f_i$ of \S\ref{sec:vlbm-model} restricted to $D=1$). Consider the first fine sub-step after a coarse collision, $t_n\to t_n+\delta t_2$, at the fine cell $R_0$ immediately across the interface from $L_0$. Specializing~\eqref{eq:1d-fv} to $R_0$,
\begin{equation}
  f_{+,R_0}^{t_n+\delta t_2} = f_{+,R_0}^{*,t_n} - \frac{\delta
  t_2}{\delta x_2}\Big(F_{+,R_0+\delta x_2/2} - F_{+,R_0-\delta
  x_2/2}\Big),
\end{equation}
and, by the same $c=\delta x/\delta t$ collapse used to reach shift streaming in \S\ref{sec:fv-basis} (now with $\delta x=\delta x_2$, $\delta t=\delta t_2$), the interior face poses no difficulty, $F_{+,R_0+\delta x_2/2}=c\,f_{+,R_0}^{*,t_n}$. The interface face $F_{+,R_0-\delta x_2/2}$ would ordinarily be filled by the post-collision value of $R_0$'s left neighbor -- and $R_0$'s left neighbor is $L_0$ -- but $L_0$ has not collided at $t_n+\delta t_2$: it collides once per coarse cycle, at $t_n$, producing $f_{+,L_0}^{*,t_n}$, and will not collide again until $t_n+\delta t_1$. Over one coarse cycle we will need $r$ such fluxes, one for each fine sub-step $t_n+k\,\delta t_2$, $k=1,\dots,r$, and only one coarse-side sample, $f_{+,L_0}^{*,t_n}$, is available to supply all of them. The question is therefore: how do we get the $r$ values $f_{+,R_0}^{t_n+k\delta t_2}$ out of the single coarse sample
$f_{+,L_0}^{*,t_n}$, together with whatever was buffered from the previous cycle?
\subsubsection{Step 3: reconstruction as a two-constraint model}
\label{sec:1d-recon}
Two coarse samples, the current $f_{+,L_0}^{*,t_n}$ and the buffered value from the previous cycle, $f_{+,L_0}^{*,t_n-\delta t_1}$, fix at most an affine model in the sub-step index $k$; we pose it in the centered form
\begin{equation}
  f_{+,R_0}^{t_n+k\delta t_2} = f^{*,t_n}_{+,L_0} + B\Big(k-\frac{r+1}{2}\Big),
  \qquad k=1,\dots,r,
  \label{eq:1d-ansatz}
\end{equation}
and fix the slope $B$ and the additive anchor from two independent requirements.\\
\emph{Consistency.} The increment per fine sub-step should reproduce, to leading order, the rate of change of $f_{+,L_0}$ resolved by the two coarse samples: the step-to-step increment of~\eqref{eq:1d-ansatz} is $B$, and matching it to the coarse-resolved rate $(f^{*,t_n}_{+,L_0}-f^{*,t_n-\delta t_1}_{+,L_0})/\delta t_1$ carried over one fine sub-step $\delta t_2=\delta t_1/r$ gives
\begin{equation}
  B = \frac{f^{*,t_n}_{+,L_0}-f^{*,t_n-\delta t_1}_{+,L_0}}{r},
  \label{eq:1d-slope}
\end{equation}
i.e. $rB=\partial_t f_+|_{L_0,t_n}+\mathcal{O}(\delta t^2)$.\\
\emph{Conservation.} By~\eqref{eq:1d-fv}, $c\,f_{+,R_0}^{t_n+k\delta t_2}$ is exactly the flux delivered into $R_0$ over the $k$-th fine sub-step, and $c\,f^{*,t_n}_{+,L_0}$ is exactly the flux $L_0$ delivers over the whole coarse cycle -- it is the only value the coarse solver ever resolves, since $L_0$ collides once per cycle. Nothing is created or destroyed at the interface, so the two totals must agree: the coarse-side throughput over $[t_n,t_n+\delta t_1]$ is $c\,f^{*,t_n}_{+,L_0}\,\delta t_1=\delta x_1\,f^{*,t_n}_{+,L_0}$, and the fine-side throughput injected at $R_0$ over the same interval is $\sum_{k=1}^r c\,f_{+,R_0}^{t_n+k\delta t_2}\,\delta t_2=\delta x_2\sum_{k=1}^r f_{+,R_0}^{t_n+k\delta t_2}$. Equating the two and using $\delta x_1=r\,\delta x_2$ gives
\begin{equation}
  \frac1r\sum_{k=1}^r f_{+,R_0}^{t_n+k\delta t_2} = f^{*,t_n}_{+,L_0}.
  \label{eq:1d-recon-mean}
\end{equation}
Because $\sum_{k=1}^r\big(k-\tfrac{r+1}{2}\big)=0$ identically, the centered ansatz satisfies~\eqref{eq:1d-recon-mean} for \emph{any} $B$ the moment the additive constant is set to $f^{*,t_n}_{+,L_0}$ -- exactly the anchor already used above. Conservation and consistency therefore pin two different, independent parts of the same ansatz: conservation fixes the constant term, consistency fixes the slope, and neither constrains the other. Substituting $B$ from~\eqref{eq:1d-slope} gives the closed form,
\begin{equation}
\begin{split}
  f_{+,R_0}^{t_n+k\delta t_2} = {}& f^{*,t_n}_{+,L_0} +
  \big(f^{*,t_n}_{+,L_0}-f^{*,t_n-\delta t_1}_{+,L_0}\big)\,
  \frac{2k-r-1}{2r}, \\
  &\qquad k=1,\dots,r,
\end{split}
  \label{eq:1d-recon}
\end{equation}
\subsubsection{Step 4: fine-to-coarse reflux}
\label{sec:1d-reflux}
After its $r$ sub-steps, $R_0$ holds $r$ genuine post-collision values of its outgoing population, $f_{-,R_0}^{*,t_n+k\delta t_2}$, $k=1,\dots,r$. Keeping only the last would discard the sub-cycle's history and, fail to conserve; instead $L_0$'s new incoming population is the plain, equally weighted average of all $r$,
\begin{equation}
  f_{-,L_0}^{t_n+\delta t_1} = \frac1r\sum_{k=1}^r
  f_{-,R_0}^{*,t_n+k\delta t_2},
  \label{eq:1d-reflux}
\end{equation}
in the spirit of the Berger--Colella reflux correction~\cite{BergerColella1989}, adapted here to the population-based bookkeeping of the vectorial lattice scheme. By the same throughput argument as Step~3, $\delta x_1 f_{-,L_0}^{t_n+\delta t_1}=\delta x_2\sum_k f_{-,R_0}^{*,t_n+k\delta t_2}$ holds identically for \emph{any} set of weights averaging to $1/r$, not only equal ones; unlike the coarse-to-fine direction, the reflux constraint carries no analog of the consistency condition on $B$ to balance against, so nothing is traded away by choosing the simplest, equal-weight average.

\subsection{Extension to 2-D}
\label{sec:2d}
On the axis-only $D2Q4$ lattice, transport is a pure per-population shift along a single Cartesian axis -- an $x$-directed population never exchanges with a $y$-directed one -- so the coupling between a
rectangular fine patch (refinement ratio $r$) and the surrounding coarse grid decomposes \emph{exactly} into four independent copies of the one-dimensional construction of \S\ref{sec:1d}, one per edge: left/right through the $x$-population pair, bottom/top through the $y$-pair. A coarse cell only diagonally adjacent to the patch, at a corner, is face-adjacent to no fine cell in any direction and needs no special treatment at all -- it simply falls outside every edge's coupling. Relative to \S\ref{sec:1d}, only two things actually change: how many cells the coarse-to-fine reconstruction has to fill, and how many values the fine-to-coarse reflux has to average.
\emph{Coarse-to-fine.} In 1-D, a coarse ring cell $L_0$ faced a single fine neighbor $R_0$; in 2-D it faces $r=\delta x_1/\delta x_2$ fine cells $R_{0,0},\dots,R_{0,r-1}$ side by side along the edge, since the fine patch shares $L_0$'s physical footprint at $r$ times the resolution in \emph{both} directions. $L_0$ still collides once per coarse cycle and still carries only the two buffered samples $f_{+x,L_0}^{*,t_n}$, $f_{+x,L_0}^{*,t_n-\delta t_1}$, but now also draws on its two tangential ring neighbors along the same edge, $L_{-1}$ and $L_{+1}$, through a slope-limited spatial correction, (Eq.~\eqref{eq:amr-slope}),
\begin{equation}
  \sigma_{+x,L_0}^{t_n} = \mathrm{minmod}\big(f^{*,t_n}_{+x,L_0}-f^{*,t_n}_{+x,L_{-1}},\ f^{*,t_n}_{+x,L_{+1}}-f^{*,t_n}_{+x,L_0}\big),
  \label{eq:2d-recon-tanslope}
\end{equation}
which vanishes whenever the two one-sided differences disagree in sign, i.e. across a genuine kink or discontinuity already resolved at the coarse level. The temporal reconstruction~\eqref{eq:1d-recon} is then handed to every one of the $r$ neighbors together with this tangential correction, evaluated at each neighbor's own offset from $L_0$'s center:
\begin{equation}
\begin{split}
  f_{+x,R_{0,j}}^{t_n+k\delta t_2} = {}& f^{*,t_n}_{+x,L_0} +
  \big(f^{*,t_n}_{+x,L_0}-f^{*,t_n-\delta t_1}_{+x,L_0}\big)\,
  \frac{2k-r-1}{2r} \\
  &+\ \sigma_{+x,L_0}^{t_n}\,\frac{2j-r+1}{2r}, \\
  &\qquad k=1,\dots,r,\ \ j=0,\dots,r-1,
\end{split}
  \label{eq:2d-recon}
\end{equation}
which recovers a spatially uniform handoff exactly whenever $\sigma_{+x,L_0}^{t_n}=0$ -- at an isolated ring cell with no same-level tangential neighbor to draw on, or wherever the limiter itself suppresses the correction near a shock. Because $\sum_{j=0}^{r-1}(2j-r+1)=0$ identically, the same telescoping cancellation already used for the temporal term in~\eqref{eq:1d-recon}, the tangential correction costs nothing conservation-wise either: averaged over $j$ at fixed $k$, Eq.~\eqref{eq:2d-recon} reduces back to the purely temporal reconstruction~\eqref{eq:1d-recon}, so the mean-preserving property~\eqref{eq:1d-recon-mean} continues to hold exactly along each row.
\emph{Fine-to-coarse.} Where 1-D reflux averaged $R_0$'s $r$ sub-step values in time, 2-D reflux must now average over the $r$ fine cells in space \emph{as well}. Over one coarse cycle the $r$ neighbors each
produce $r$ genuine post-collision values -- $r^2$ in total, one per (fine-cell, sub-step) pair -- and $L_0$'s new incoming population is their plain average over both indices,
\begin{equation}
  f_{-,L_0}^{t_n+\delta t_1} = \frac1{r^2}
  \sum_{k=1}^r\sum_{j=0}^{r-1} f_{-,R_{0,j}}^{*,t_n+k\delta t_2},
  \label{eq:2d-reflux}
\end{equation}
which reduces to~\eqref{eq:1d-reflux} exactly at $r=1$. Applied independently, with the appropriate population pair, on each of the four patch edges, Eqs.~\eqref{eq:2d-recon}--\eqref{eq:2d-reflux} conserve mass, momentum, and energy exactly.
\subsection{Adaptive grid refinement}
\label{sec:amr}
\begin{figure}[t]
\centering
\begin{tikzpicture}[
  node distance=7mm and 10mm,
  every node/.style={font=\footnotesize},
  box/.style={draw,thick,rounded corners,align=center,text width=4.0cm,minimum height=9mm,inner sep=2.5mm,fill=blue!6},
  decbox/.style={draw,thick,align=center,text width=4.0cm,minimum height=9mm,inner sep=2.5mm,fill=orange!8},
  arr/.style={-{Latex[length=2mm]},thick}
]
 \node[box] (sensor) {evaluate sensor $\hat s(\bm x)$ at every cell, Eq.~\eqref{eq:amr-sigma}};
 \node[decbox,below=of sensor] (tag) {tag cells above tolerance; dilate by buffer; form new window(s)};
 \node[box,below=of tag] (regrid) {regrid: merge exiting cells (Eq.~\eqref{eq:amr-merge}), then split entering cells (Eq.~\eqref{eq:amr-split}) -- finest level first};
 \node[box,below=of regrid] (evolve) {advance $n_{\rm regrid}$ coarse cycles; reconstruction/reflux Eqs.~\eqref{eq:1d-recon}--\eqref{eq:2d-reflux} unmodified at whichever interfaces are currently active};
 \draw[arr] (sensor) -- (tag);
 \draw[arr] (tag) -- (regrid);
 \draw[arr] (regrid) -- (evolve);
 \draw[arr] (evolve.east) -- ++(1.1,0) |- (sensor.east);
\end{tikzpicture}
\caption{The adaptive grid refinement cycle.}
\label{fig:amr-flowchart}
\end{figure}
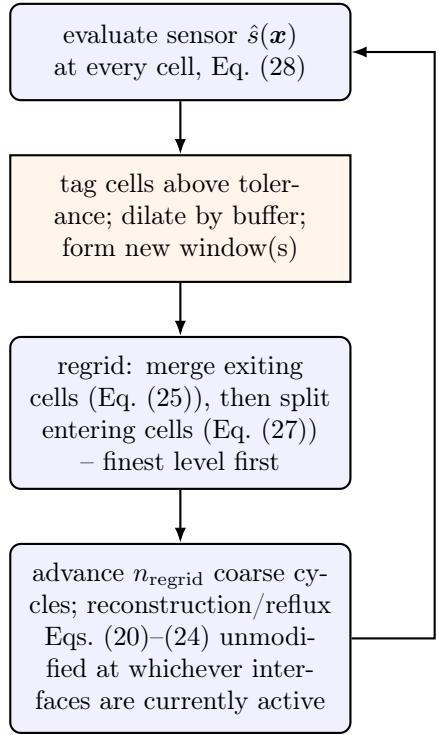

The next natural step for mesh refinement, is extension to adaptive mesh refinement. We now discuss making refinement zones a function of the solution, while changing nothing about the coupling that ties one level to the next: adaptivity only ever decides, and periodically revises, \emph{which} interfaces exist; the reconstruction and reflux operators themselves are applied completely unmodified at whichever interfaces happen to be active on a given cycle.\\
Making refinement zones time-dependent raises one question with no analog in \S\ref{sec:1d}--\S\ref{sec:2d}: whenever a cell's refinement status changes, its populations must be produced on the new grid from data that only exists on the old one. Going from coarse to fine we call this operation \emph{splitting}; its reverse, from fine to coarse, we call \emph{merging}. Both must conserve exactly, for the same reason the reconstruction and reflux operators do: any cell entering or leaving a refined region should add or remove nothing from the running totals of mass, momentum, and energy.\\
Merging has no freedom to speak of: the unique conservative reduction of $r^D$ fine values to one coarse value is their plain average,
\begin{equation}
  \bm f_i = \frac1{r^D}\sum_{j\in i} \bm f_j,
  \label{eq:amr-merge}
\end{equation}
for every fine cell $j$ nested inside coarse cell $i$. Splitting is the reverse: splitting must instead distribute a single coarse value among $r^D$ children, and this is where a genuine choice appears. The simplest option, a flat copy $\bm f_j=\bm f_i$ for every child, is conservative for any slope -- trivially, since it carries none -- but it discards whatever spatial variation is already resolved on the coarse side, exactly the kind of stairstep artifact grid refinement is meant to avoid. We instead split with a minmod-limited, tensor-product \emph{linear} reconstruction, one independent slope per Cartesian direction $\alpha=1,\dots,D$,
\begin{equation}
  \sigma_{i,\alpha} = \mathrm{minmod}\big(\bm f_i-\bm f_{i-\hat\alpha},\ \bm f_{i+\hat\alpha}-\bm f_i\big),
  \label{eq:amr-slope}
\end{equation}
built from $i$'s immediate coarse neighbors along $\alpha$ exactly as in any standard MUSCL-type limiter, and assign each child $j$ (indexed along axis $\alpha$ by $k_\alpha=0,\dots,r-1$, at signed offset $\xi_{k_\alpha}=\tfrac{2k_\alpha-r+1}{2r}\in(-\tfrac12,\tfrac12)$ from $i$'s center) the value
\begin{equation}
  \bm f_j = \bm f_i + \sum_{\alpha=1}^D \sigma_{i,\alpha}\,\xi_{k_\alpha}.
  \label{eq:amr-split}
\end{equation}
Because the offsets $\xi_{k_\alpha}$ sum to zero over $k_\alpha=0,\dots,r-1$, averaging~\eqref{eq:amr-split} over all $r^D$ children reproduces $\bm f_i$ exactly for \emph{any} choice of slope -- conservation is guaranteed independent of the limiter -- while the minmod slope itself supplies a genuine, monotonicity-preserving sub-cell profile in the smooth or weakly varying case and collapses back to the flat copy at a true discontinuity, where the limiter switches off. Splitting and merging are each applied population-by-population, identically to every one of the $\bm f_i$, so both operators inherit the same exact conservation as the ordinary reconstruction and reflux of \S\ref{sec:1d}--\S\ref{sec:2d}.\\
In addition to these two operators, a practical implementation needs one further ingredient: a sensor, or tagging criterion, that decides where refinement is actually warranted at a given instant. Choosing and tuning such a sensor is a substantial topic on its own and is not the focus of the present study; we defer a more detailed discussion of refinement criteria and sensors to~\cite{Strassle2026Sensors}. Here we reuse, essentially unchanged, the per-field flux-jump estimate $\hat s_k=\delta x\,(\nabla\cdot\bm Q)_k/\max_{\bm x,\alpha}|Q_{\alpha,k}|$ already introduced for the dissipation sensor of Eq.~\eqref{eq:sensor-B}, but combine the $D+2$ components with a max norm rather than the root-mean-square used there,
\begin{equation}
  \hat s(\bm x) = \max_k |\hat s_k(\bm x)|,
  \label{eq:amr-sigma}
\end{equation}
so that a feature breaking only a single conserved field is not diluted by the others; a cell is tagged for refinement wherever $\hat s$ exceeds a fixed tolerance, and the tagged set is dilated by a small buffer of cells before being converted into a rectangular footprint per level, so that the coarse-fine interface always trails the feature it is following rather than clipping it. Figure~\ref{fig:amr-flowchart} summarizes how the two operators and the sensor fit together into a single adaptive cycle: sensor evaluation and regridding (merge, then split, finest level first) happen only once every $n_{\rm regrid}$ coarse steps, with ordinary sub-cycled time evolution -- entirely as described in \S\ref{sec:1d}--\S\ref{sec:2d} -- filling the steps in between.
\section{Numerical validation}
\label{sec:validation}
\subsection{Conservation and convergence}
\label{sec:val-conservation-convergence}
We open with three canonical, minimal-topology checks -- a single refinement patch or interface, never more than one at a time -- that isolate the coupling's two most basic guarantees in the cleanest setting available for each: two one-dimensional problems, one smooth and linear and one genuinely nonlinear, for which exact conservation can be checked directly against a known analytic answer (parts~(a) and~(b)), and a fully smooth, exact solution for which a formal convergence \emph{rate} can actually be measured (part~(c)).
\subsubsection{One-dimensional acoustic pulse across the interface}
\label{sec:val-acoustic}
We first look into the effect of the refinement of on the propagation speed of eigen-modes, more specifically the normal eigen-modes, i.e. acoustic waves. A single interface at $x=0$ separates a coarse region $x\in[-1,0]$ ($\delta x_1=0.01$) from a fine region $x\in[0,1.5]$ ($\delta x_2=\delta x_1/r=0.005$, $r=2$), coupled by Eqs.~\eqref{eq:1d-recon}--\eqref{eq:1d-reflux} exactly as derived in \S\ref{sec:1d}, with $\beta=0.99$ (chosen deliberately close to $1$ so that numerical dissipation contributes negligibly and the measurement isolates the coupling's own effect). The ambient state is uniform, $\rho_0=1$, $u_0=0$, with $p_0$ chosen so the ambient sound speed is exactly $c_{s0}=\sqrt{\gamma p_0/\rho_0}=1$. A small-amplitude ($A=0.01\,\rho_0$), purely right-moving linear acoustic pulse is imposed as the initial condition, using the linearized right-moving Riemann-invariant relation to avoid exciting a left-moving companion wave,
\begin{equation}
\begin{aligned}
  \rho(x) &= \rho_0 + A\,e^{-(x-x_0)^2/2\sigma^2}, \\
  u(x) &= \frac{c_{s0}}{\rho_0}\big(\rho(x)-\rho_0\big), \qquad
  p(x) = p_0 + c_{s0}^2\big(\rho(x)-\rho_0\big),
\end{aligned}
  \label{eq:val-acoustic-ic}
\end{equation}
with $\sigma=0.05$ and the pulse centered at $x_0=-0.6$, comfortably inside the coarse region.

At every coarse sub-step, the pulse's peak position is located to sub-grid accuracy by a three-point parabolic fit to the density perturbation $\rho(x,t)-\rho_0$ around its discrete maximum, evaluated directly on the solver's own combined, non-uniformly spaced grid -- so the measurement makes no separate assumption about which side of the interface the peak currently occupies. The resulting
trajectory $x_{\rm peak}(t)$ is converted to an effective sound speed $c_{\rm eff}(t)$ by a local (sliding-window) linear regression.
\begin{figure}[h!]
\centering
\includegraphics[width=0.9\linewidth]{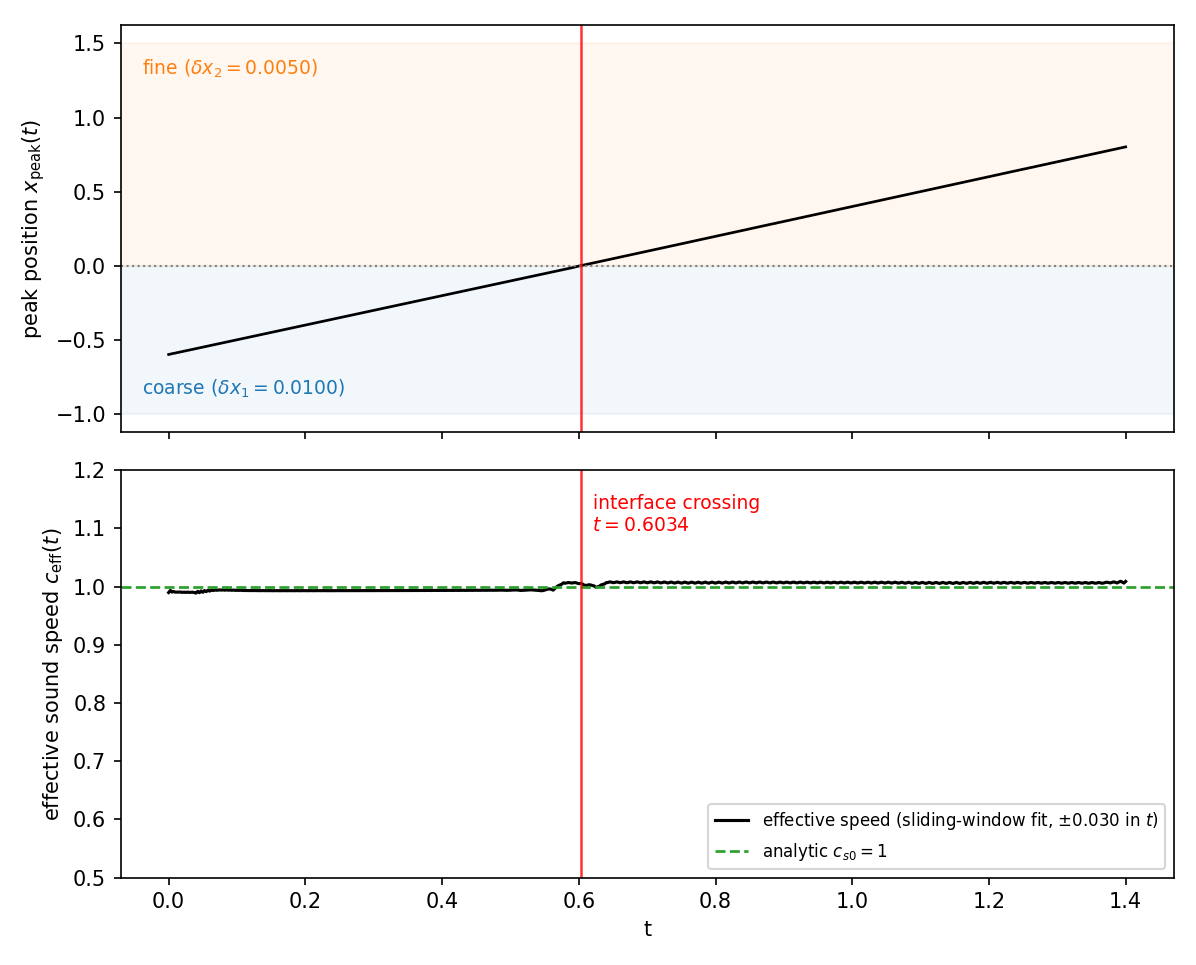}
\caption{One-dimensional acoustic pulse crossing the refinement interface. Top: sub-grid peak position $x_{\rm peak}(t)$, shaded by region (coarse below $x=0$, fine above). Bottom: effective sound speed
$c_{\rm eff}(t)$ against the analytic $c_{s0}=1$ (dashed), on a deliberately wide axis.}
\label{fig:val-acoustic}
\end{figure}
Figure~\ref{fig:val-acoustic} shows the pulse crossing the interface with no visible reflection, dissipation spike, or drift in its trajectory. The measured crossing time, $t_{\rm cross}=0.6034$, agrees with the exact geometric prediction $|x_0|/ c_{s0}=0.600$ to within $0.6\%$. Away from the crossing itself, $c_{\rm eff}$ sits about $0.7\%$ below $c_{s0}$ on the coarse side and about $0.7\%$ above it on the fine side; the relative error in $c_{\rm eff}$ stays below $1.2\%$ throughout the run. The coupling introduces no speed error, dissipation, or reflection of its own beyond the local truncation, and the interface is acoustically transparent.

\subsubsection{Sod shock tube across the interface}
\label{sec:val-sod}
\begin{figure}[h!]
\centering
\includegraphics[width=0.7\linewidth]{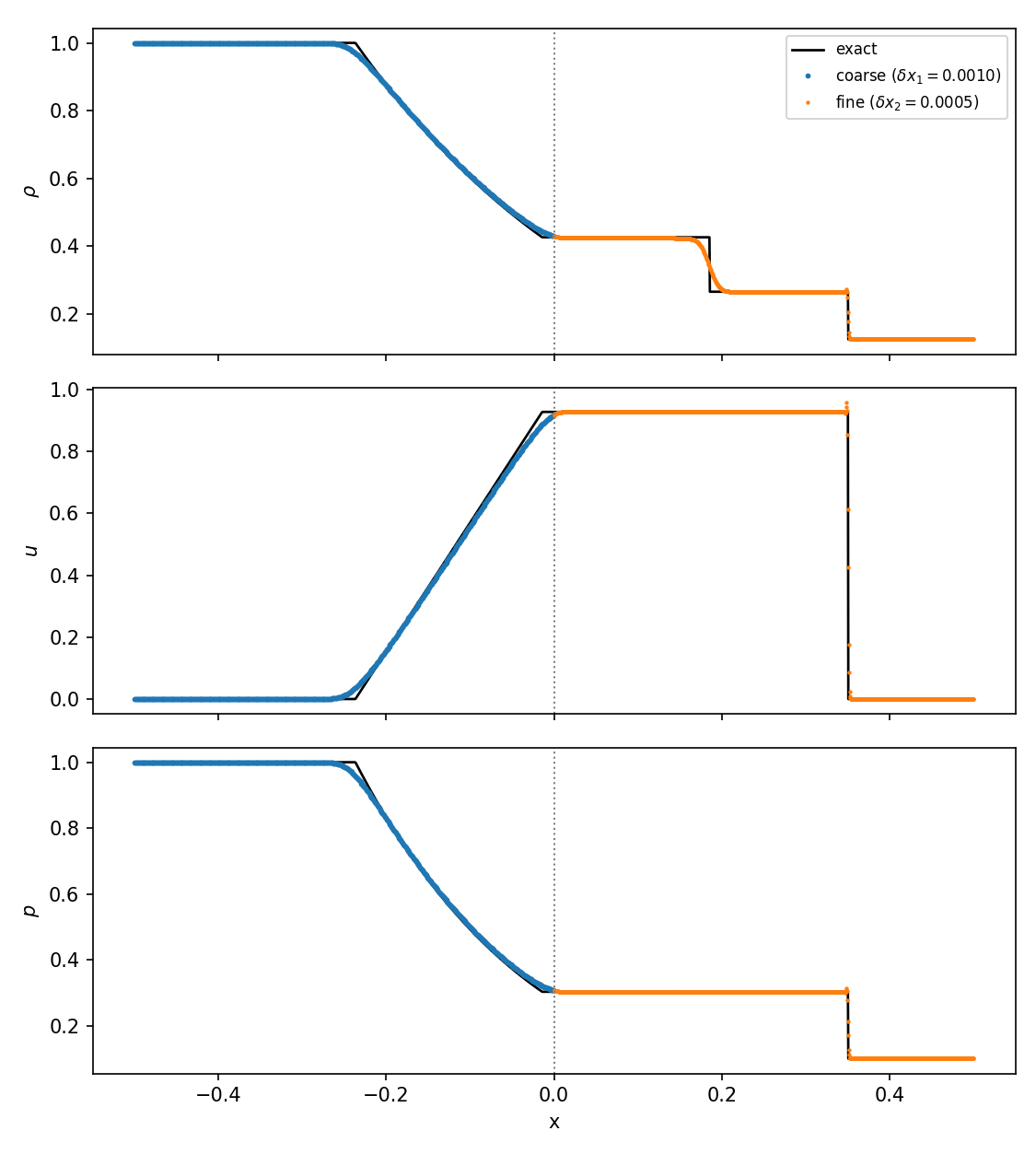}
\caption{Sod shock tube with the refinement interface at the domain midpoint (dotted line), $t=0.2$: density, velocity, and pressure against the exact Riemann solution~\cite{Toro2009}.}
\label{fig:val-sod-profiles}
\end{figure}

Having checked wave-speed fidelity for a small, smooth disturbance in the previous section, we now consider a problem with a shock, a contact discontinuity, and a rarefaction fan, all of which cross the refinement interface, using the classic Sod shock tube~\cite{Toro2009}. The interface is placed exactly at the domain midpoint, coinciding with the initial diaphragm: $x\in[-0.5,0]$ is coarse ($\delta x_1=0.001$), $x\in[0,0.5]$ is fine ($\delta x_2=\delta x_1/r=0.0005$, $r=2$), with initial data $(\rho,u,p)_L=(1,0,1)$ for $x<0$ and $(\rho,u,p)_R=(0.125,0,0.1)$ for $x>0$, $\gamma=1.4$, $\beta=0.9$, run to $t_{\rm end}=0.2$. Figure~\ref{fig:val-sod-profiles} shows excellent agreement with the exact solution on both sides of the interface, with the rarefaction fan resolved on the coarse grid and the contact and shock -- by $t=0.2$, both located within the fine region -- captured sharply and without any visible trace of having crossed a resolution change earlier in the run.
\begin{figure}[h!]
\centering
\includegraphics[width=0.8\linewidth]{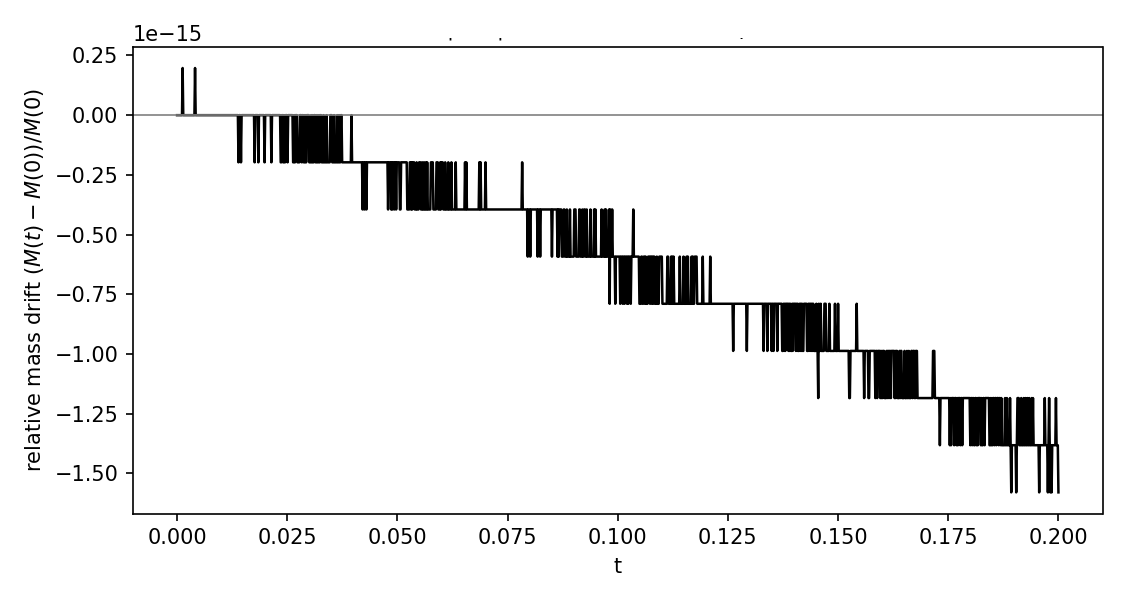}
\caption{Total mass $M(t)=\delta x_1\sum_{\rm coarse}\rho+\delta x_2\sum_{\rm fine}\rho$ through the same run, relative to $M(0)$. The
vertical scale is $10^{-15}$.}
\label{fig:val-sod-mass}
\end{figure}
Figure~\ref{fig:val-sod-mass} tracks the total mass over the whole run: the relative drift never exceeds $\sim10^{-15}$, i.e.\ floating-point roundoff, confirming that the exact-conservation argument of \S\ref{sec:1d-recon}--\ref{sec:1d-reflux} holds here.
\subsubsection{Stationary isentropic vortex: grid convergence}
\label{sec:val-vortex-convergence}
We next look into a configuration to probe the effect of the refinement interface on the convergence behavior of the solver. We use the Yee/Shu isentropic vortex construction: a purely azimuthal velocity field $u_\phi(r)=\frac{\beta_v}{2\pi R_c}\,r\,e^{\frac12(1-r^2/R_c^2)}$ around a center $\bm x_c$, closed by imposing radial momentum balance $dp/dr=\rho\,u_\phi(r)^2/r$ along the ambient isentrope. For an ideal gas ($\gamma=1.4$) this closes in the classical form
\begin{equation}
\begin{aligned}
\Delta T(r) &= -\frac{(\gamma-1)\beta_v^2}{8\gamma\pi^2}\,e^{1-r^2/R_c^2}, \\
T(r) &= T_\infty + \Delta T(r), \quad
\rho(r) = \rho_\infty\Big(\frac{T(r)}{T_\infty}\Big)^{\!\frac{1}{\gamma-1}}, \\
p(r) &= \rho(r)\,T(r),
\end{aligned}
\label{eq:vortex-exact}
\end{equation}
on the periodic domain $[0,1]^2$, with $\rho_\infty=p_\infty=1$, $\bm x_c=(0.5,0.5)$, circulation $\beta_v=2$, and core radius $R_c=0.05$: small enough relative to the domain that the profile is negligible ($<10^{-4}$ relative amplitude) already at the boundary of a centered patch spanning half the domain, and negligible ($\sim10^{-20}$) at the periodic edge itself, so the periodic wrap introduces no inconsistency. The resulting vortex peaks at $|\bm u|_{\max}=\beta_v/(2\pi)\approx0.318$ (Mach $\approx0.27$) at $r=R_c$, with a mild $\sim\!1.4\%$ density/pressure dip at the core.\\
We run with acoustic scaling ($\delta t=\delta x/c$, $c$ fixed from the initial data. A base grid
$N_0\in\{32,64,128,256\}$ carries, in the refined series, a single centered patch at $r=2$ over $[0.25,0.75]^2$ run to $t_{\rm end}=0.5$ at each of the four base resolutions, alongside a matched unrefined run at the same four resolutions with identical solver settings.
\begin{figure}[h!]
\centering
\includegraphics[width=0.7\linewidth]{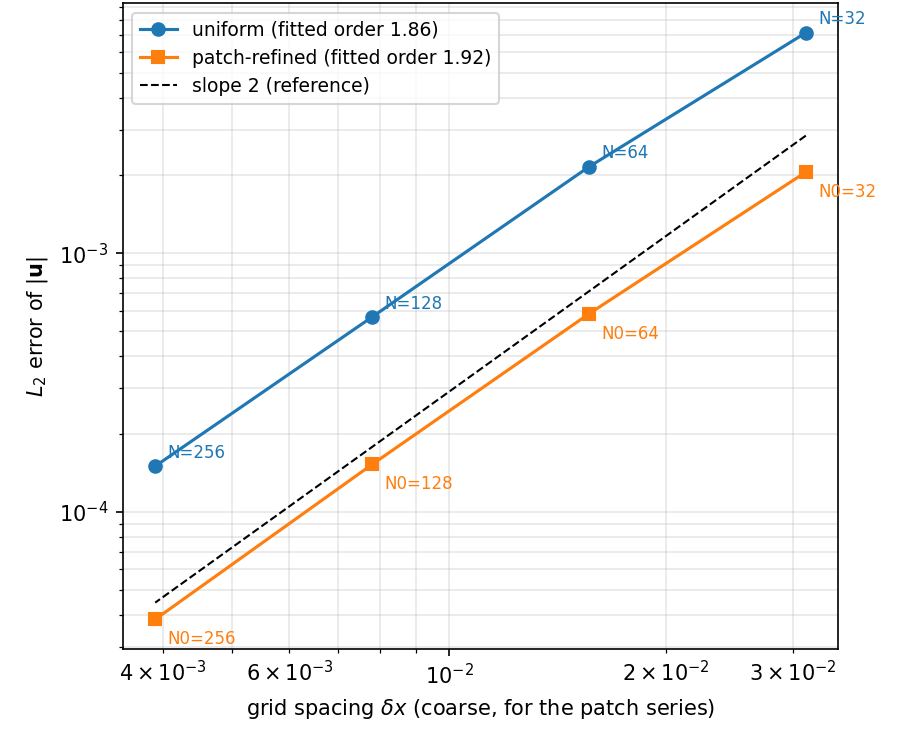}
\caption{Stationary isentropic vortex, $t_{\rm end}=0.5$: $L_2$ error of $|\bm u|$ vs.\ (coarse) grid spacing $\delta x$, log--log, for the unrefined sequence and the patch-refined sequence at the same four base resolutions $N_0=32,64,128,256$. A slope-2 reference line is shown for
comparison.}
\label{fig:vortex-convergence}
\end{figure}
Figure~\ref{fig:vortex-convergence} shows the resulting $L_2$ error of the velocity magnitude against the exact solution~\eqref{eq:vortex-exact}, evaluated over the whole domain at each resolution. Both series converge close to second order and show no sign of an order reduction due to the refinement
interface. This confirms directly that the coupling introduces no accuracy degradation.
\begin{figure}[h!]
\centering
\includegraphics[width=0.9\linewidth]{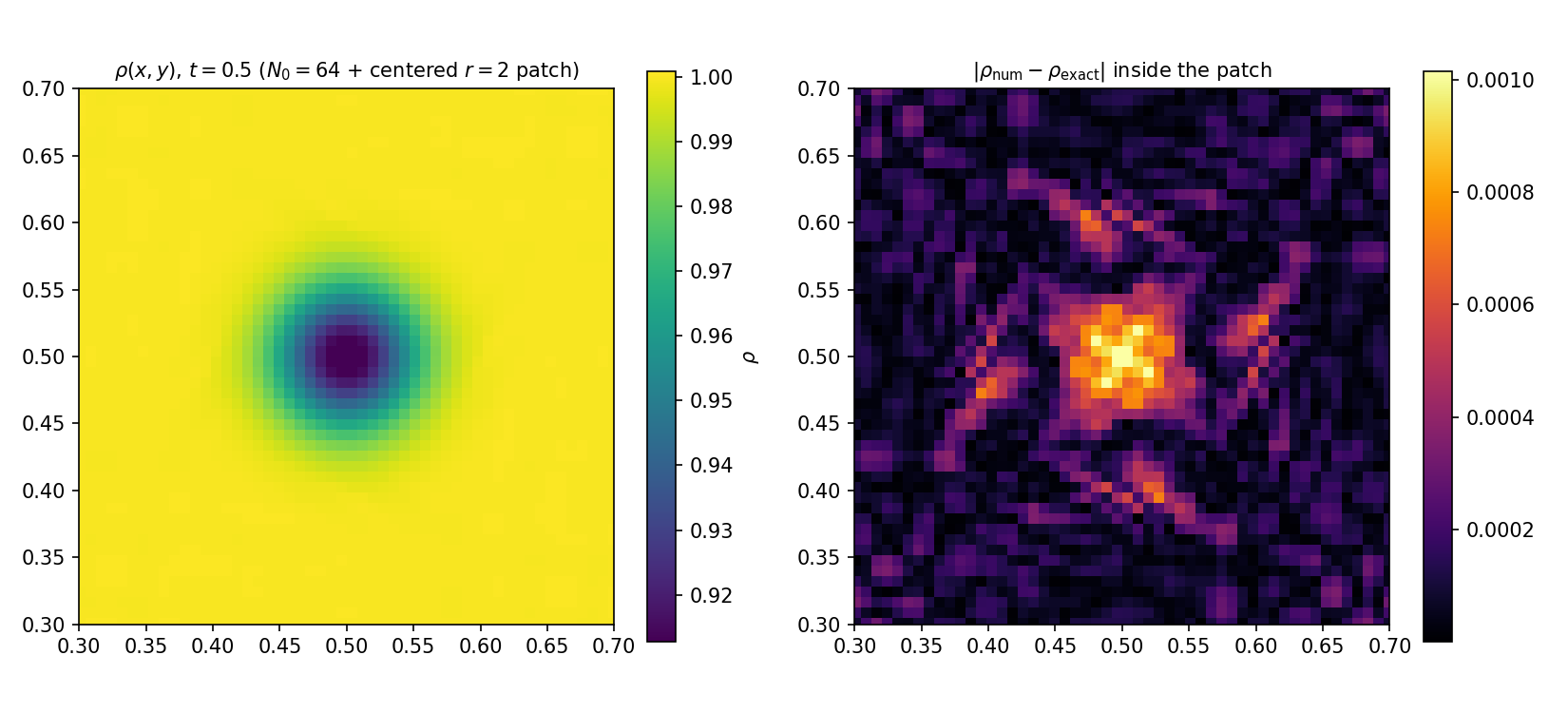}
\caption{Stationary isentropic vortex, $t=0.5$, $N_0=64$ plus a centered $r=2$ patch (dashed outline): density field, zoomed to the patch footprint (left), and the pointwise density error against the exact
solution inside the patch (right).}
\label{fig:vortex-field}
\end{figure}
Figure~\ref{fig:vortex-field} shows the density field and its pointwise error at $N_0=64$: the error is concentrated at the vortex core itself, where the profile's curvature is largest, with no visible artifact along the patch boundary -- consistent with the coupling contributing no error of its own beyond the smooth truncation error already present in the underlying scheme.
\begin{figure}[h!]
\centering
\includegraphics[width=\linewidth]{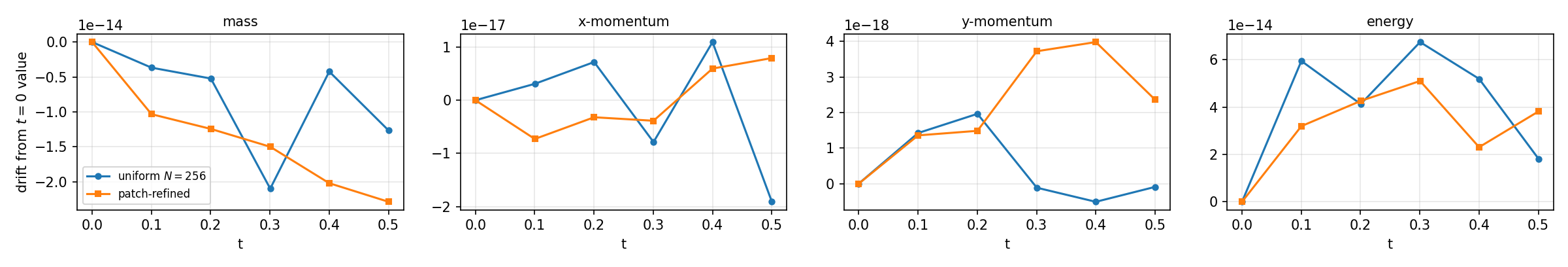}
\caption{Exact conservation under periodic boundaries, $N_0=256$: total mass, $x$- and $y$-momentum, and energy, relative to their $t=0$ values, at six instants up to $t=0.5$.}
\label{fig:vortex-conservation}
\end{figure}
Because this case is periodic, \emph{every} conserved quantity integrated over the domain must be exactly constant in time. Figure~\ref{fig:vortex-conservation} confirms this at $N_0=256$: total mass, both momentum components, and energy all stay within $\sim10^{-14}$ of their initial value throughout the run, for both the unrefined and patch-refined series.
\subsection{Explosion in a box}
\label{sec:val-explosion}
The next configuration we consider is the explosion-in-a-box configuration, also used in~\cite{Amroc,PhysRevE106015301}. On a $[0,1]^2$ domain with $\gamma=1.4$, a circular region of elevated density and pressure, both initially at rest, is embedded in an ambient state also at rest,
\begin{equation}
(\rho,u,v,p) =
\begin{cases}
(5,0,0,5), & |(x,y)-(0.4,0.4)| < 0.3,\\
(1,0,0,1), & \text{otherwise},
\end{cases}
\label{eq:explosion-ic}
\end{equation}
with reflective boundary conditions on all four edges of the box. The circular front expands, reflects off each wall in turn, and the reflected waves interact in an increasingly complex pattern.\\
A base grid $N_0=256$ carries a single centered refinement patch at $r=2$ ($\delta x_1=1/512$) over
$[0.25,0.75]^2$ -- the same centered, single-rectangle topology and footprint as the vortex of \S\ref{sec:val-vortex-convergence}, now combined with reflective rather than periodic domain edges.
\begin{figure}[h!]
\centering
\includegraphics[width=0.9\linewidth]{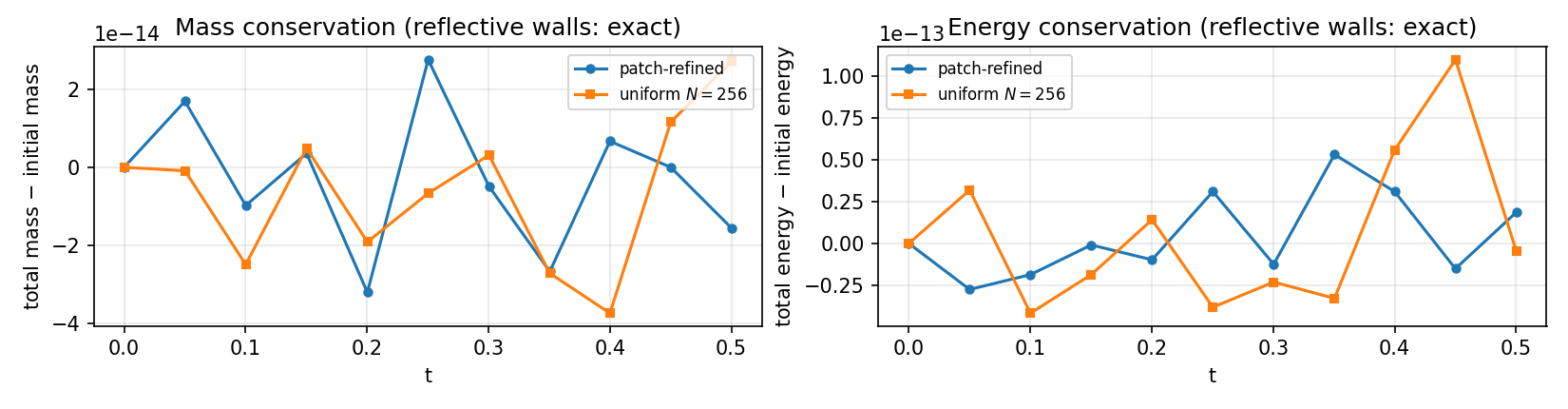}
\caption{Explosion in a box: total mass and energy (patch-refined and uniform $N=256$ runs), relative to their $t=0$ values, sampled every $\Delta t=0.05$ up to $t=0.5$.}
\label{fig:val-explosion-conservation}
\end{figure}
Figure~\ref{fig:val-explosion-conservation} tracks total mass and energy (summed over the coarse grid outside the patch hole plus the fine patch) at ten instants up to $t=0.5$, for both the patch-refined run and a matched uniform-$N{=}256$ baseline. The maximum drift over the entire run is $3.2\times10^{-14}$ in mass and $5.3\times10^{-14}$ in energy for the patch-refined run, and $3.7\times10^{-14}$ and $1.1\times10^{-13}$ respectively for the uniform baseline.
\begin{figure}[h!]
\centering
\includegraphics[width=0.6\linewidth]{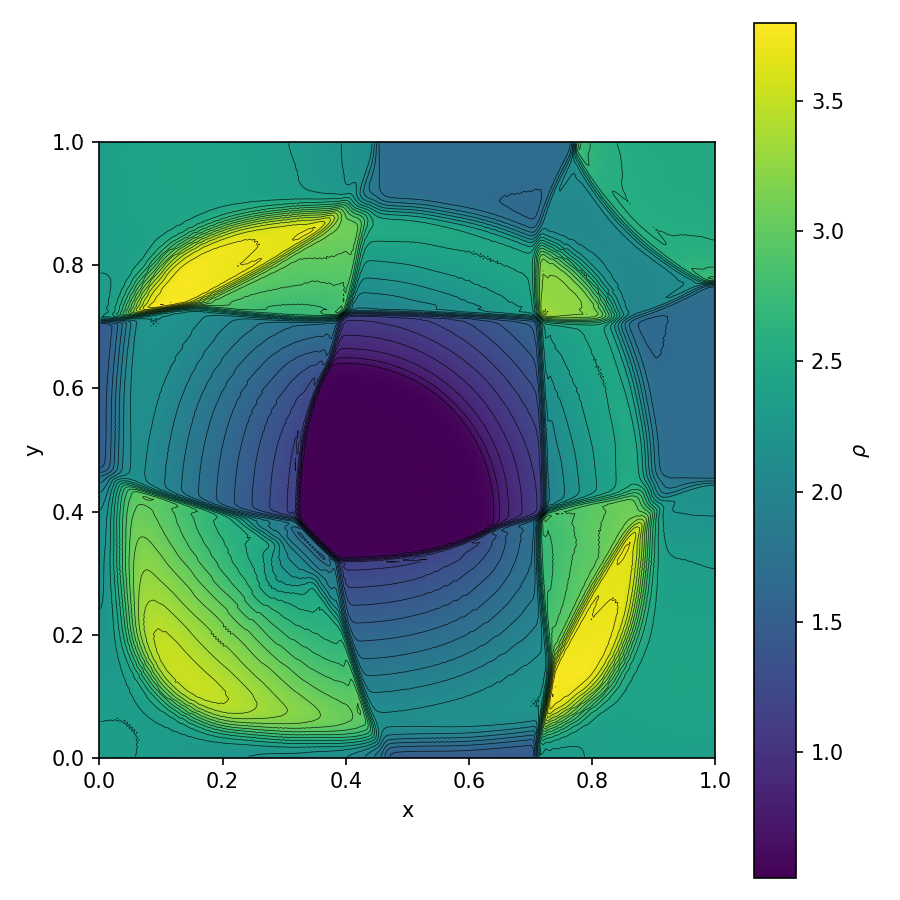}
\caption{Explosion in a box, uniform $N=256$, $t=0.5$: density field with 30 equally spaced contours over $\rho\in[0.52,3.8]$.}
\label{fig:val-explosion-field}
\end{figure}
Figure~\ref{fig:val-explosion-field} shows the density field at $t=0.5$ from a uniform run at the reference paper's own resolution, $N=256$, contoured on the same $\rho\in[0.52,3.8]$ scale used: a low-density band structure running along both diagonals, four high-density pockets near the corners where
reflected fronts pile up, and a rarefied core left behind near the domain's center. This qualitative pattern, and the final density range it spans ($\rho\in[0.530,3.802]$ here against the reported
$[0.52,3.8]$), agree closely with the published result.
\begin{figure}[h!]
\centering
\includegraphics[width=0.6\linewidth]{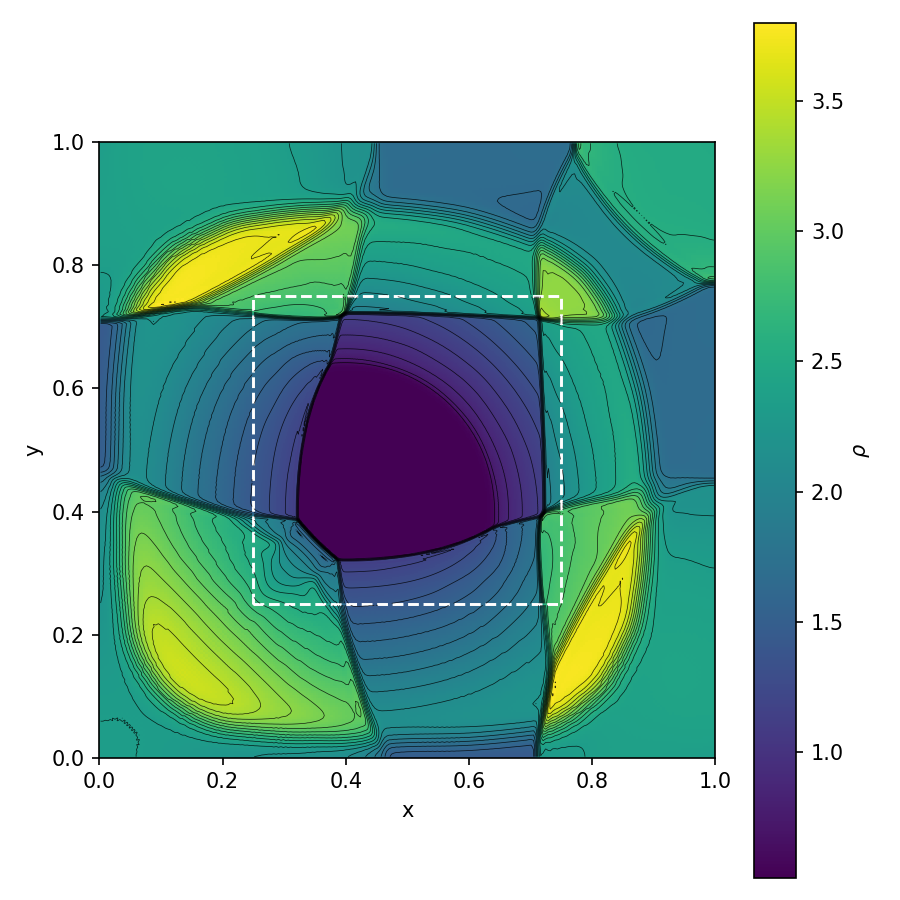}
\caption{Explosion in a box, patch-refined ($N_0=256$ plus a centered
$r=2$ patch, dashed outline), $t=0.5$. The field is visually
continuous across the patch boundary, with no artifact from the
resolution jump or from the reflective walls, which lie well outside
the patch and are handled entirely on the coarse grid.}
\label{fig:val-explosion-patch}
\end{figure}
\begin{figure}[h!]
\centering
\includegraphics[width=\linewidth]{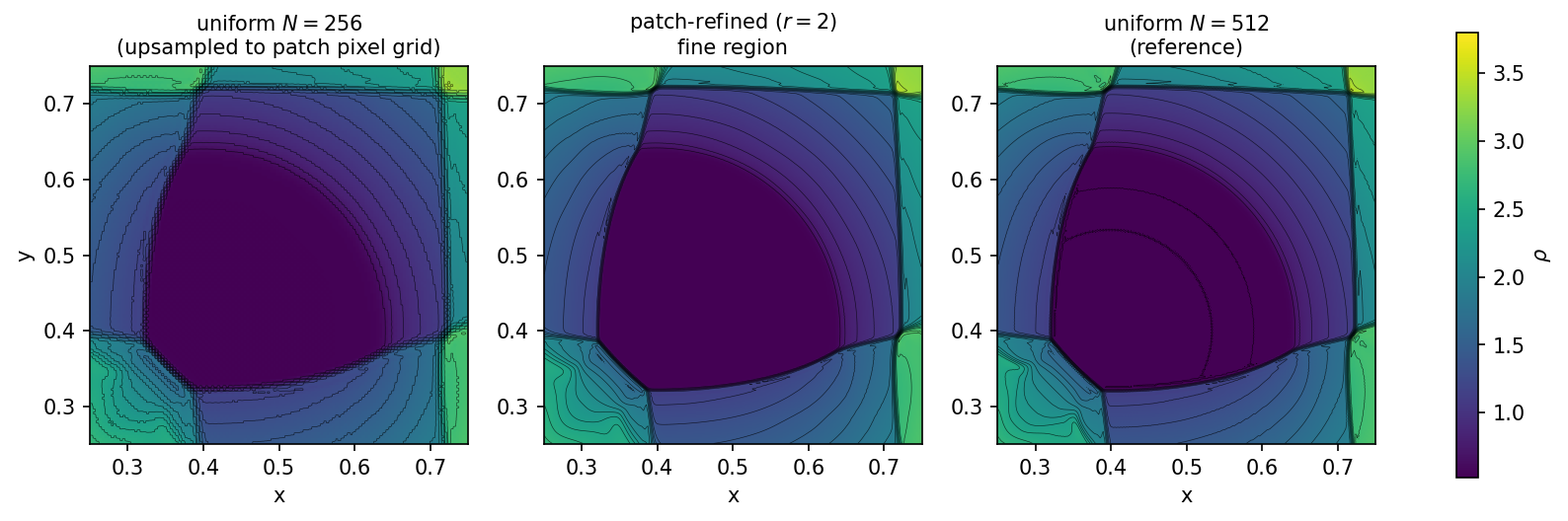}
\caption{Explosion in a box, $t=0.5$, identical solver settings, varying only grid/refinement, restricted to the patch footprint $[0.25,0.75]^2$: uniform $N=256$ upsampled onto the patch's pixel grid (left), the patch-refined run's own fine region (center), and an
independent uniform $N=512$ reference (right).}
\label{fig:val-explosion-isolated}
\end{figure}
Figure~\ref{fig:val-explosion-patch} shows the same field from the patch-refined run: the density is
continuous across the patch boundary, with no visible artifact from either the resolution jump or the reflective walls. We compare the patch-refined fine region against an independent, uniform $N=512$
reference run restricted to the patch footprint (Fig.~\ref{fig:val-explosion-isolated}): the $L_1$ density error drops from $0.0261$ for the upsampled uniform-256 field to $0.0102$ for the patch-refined field, a $2.6\times$ reduction.
\subsection{2-D Riemann cases}
\label{sec:val-2d-riemann}
As a final case, to illustrate the robustness of the approach we consider configurations 3 and 12 of the Lax--Liu classification.
\subsubsection{Multi-patch, three-level refinement: Configuration 3}
\label{sec:val-case3-cross}
\begin{figure}[h!]
\centering
\includegraphics[width=0.8\linewidth]{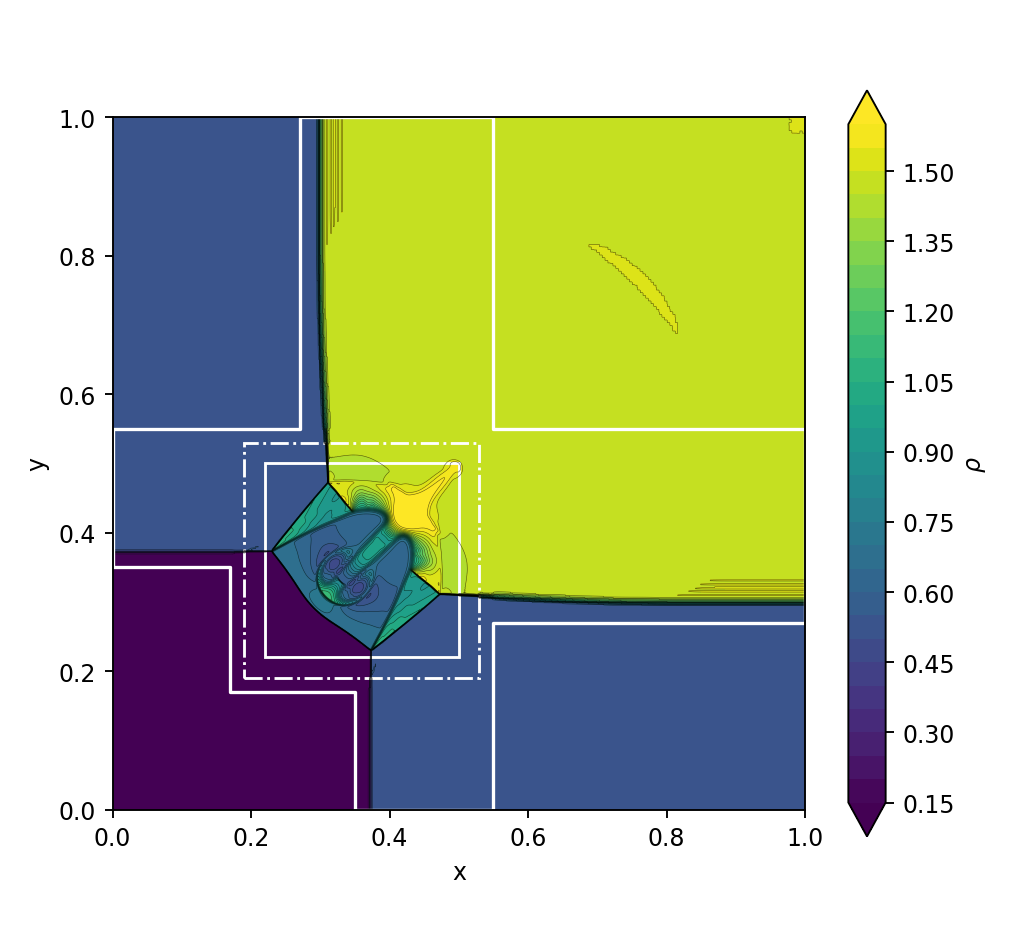}
\caption{Configuration~3, cross-shaped three-level refinement, $t=0.3$ (355 level-0 steps): density field and isocontours over the full domain, composited onto a single grid (finest resolution available at each point) before contouring, with the level-1, level-2, and level-3 footprints outlined in white.}
\label{fig:val-case3cross-field}
\end{figure}

We first consider Configuration~3 of the Lax--Liu classification of two-dimensional Riemann
problems~\cite{LaxLiu1998,KurganovTadmor2002}, in which all four quadrant-boundary edges are shocks, producing a diamond-shaped shock--shock interaction region and a central Kelvin--Helmholtz mushroom
jet, run to $t_{\rm end}=0.3$. A base grid $N_0=256$ ($\delta x_0=1/256$) carries three nested levels of refinement, cross-shaped to follow the four outgoing shock branches while concentrating the finest resolution on the central interaction: level~1 at $r_1=2$ ($\delta x_1=\delta x_0/2$), level~2 at $r_2=4$ relative to level~0 ($\delta x_2=\delta x_0/4$), nested inside level~1 around the interaction region, and level~3 at $r_3=8$ relative to level~0 ($\delta x_3=\delta x_0/8$), nested inside level~2 around the roll-up itself. Figure~\ref{fig:val-case3cross-field} shows the resulting topology along with the obtained density field.
We isolate the effect of the refinement by comparing three runs sharing identical solver settings and differing only in grid/refinement: a uniform grid at $N=256$ (the base resolution, 328 steps), the cross-shaped, three-level refined run (355 level-0 steps), and a uniform grid at $N=1024$ (matching the level-2 resolution, 1376 steps).
\begin{figure}[h!]
\centering
\includegraphics[width=\linewidth]{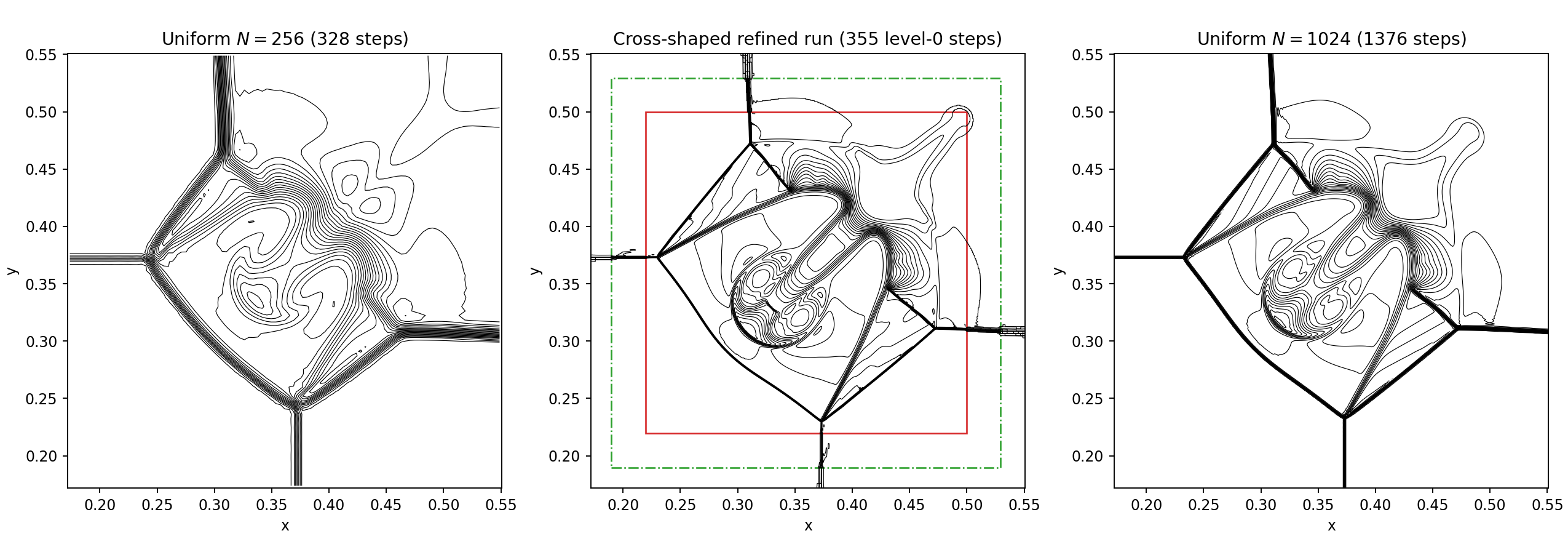}
\caption{Configuration~3, $t=0.3$, identical solver settings, varying only grid/refinement, zoomed on the shock--shock interaction region: uniform $N=256$ (left), the cross-shaped refined run (center, patch boundaries overlaid), and uniform $N=1024$ (right), density with
matched isocontours (30 equally spaced levels, $\rho\in[0.15,1.6]$, the
same evenly-spaced-contour convention used by Lax and Liu's own Fig.~3~\cite{LaxLiu1998}.}
\label{fig:val-case3cross-isolated}
\end{figure}
Figure~\ref{fig:val-case3cross-isolated} shows the result. The $L_1$ density error against that reference is $0.0048$ for the cross-shaped refined run, against $0.0456$ for the uniform-$N{=}256$ run -- a $9.6\times$ reduction, confirming that the accuracy gain is attributable to the refinement itself under matched dissipation and time-stepping.
\subsubsection{Symmetric multi-patch benchmark: Configuration 12}
\label{sec:val-case12}
\begin{figure}[h!]
\centering
\includegraphics[width=0.6\linewidth]{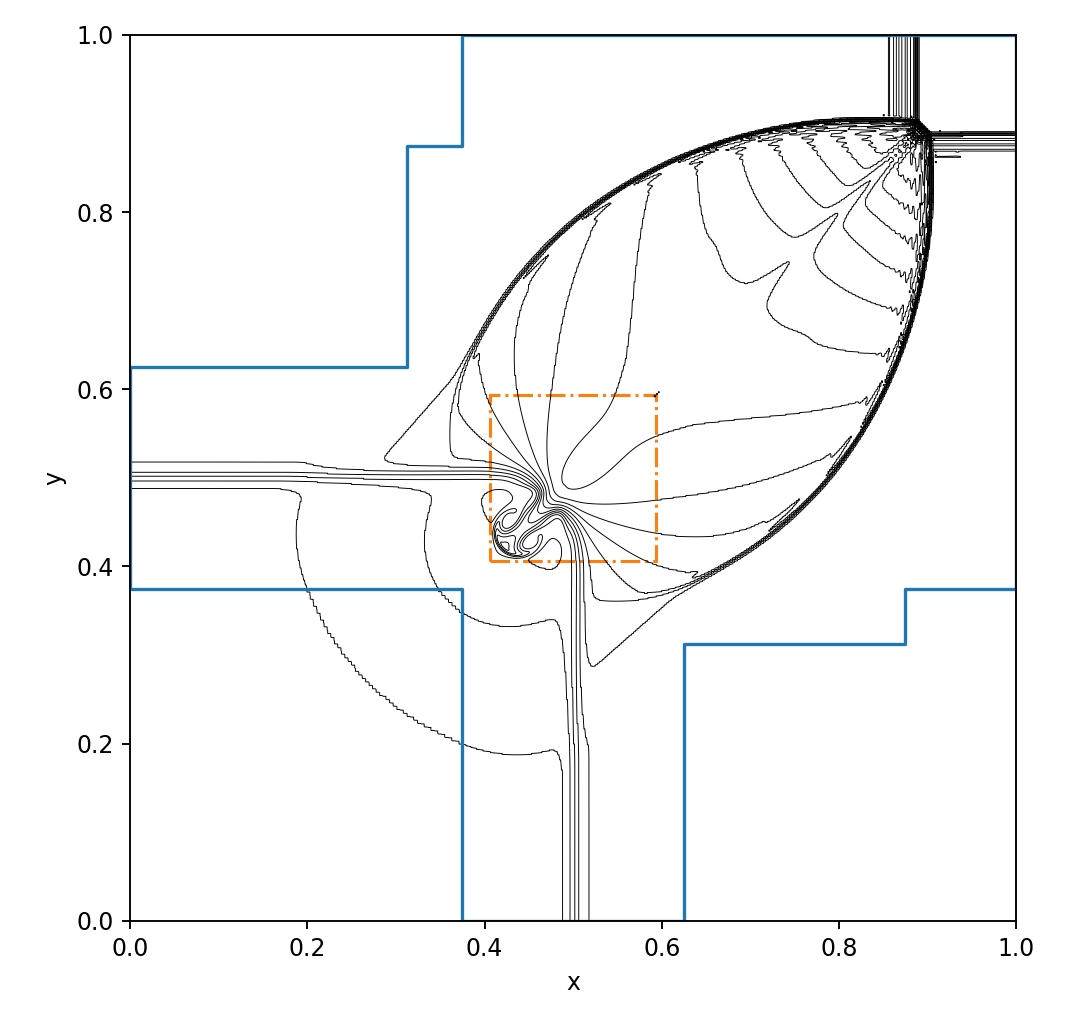}
\caption{Configuration~12, $t=0.25$: density isocontours of the
two-level staircase run (323 level-0 steps), composited onto a single
grid (finest resolution available at each point) before contouring,
with the level-1 (blue) and level-2 (orange) footprints outlined.}
\label{fig:val-case12-field}
\end{figure}

Next we consider Configuration~12 of the Lax--Liu classification on $[0,1]^2$ with discontinuities at $x=0.5$, $y=0.5$, $\gamma=1.4$,
\begin{equation}
\begin{aligned}
  ({\rm NE})&:\ (\rho,u,v,p)=(0.5313,\,0,\,0,\,0.4), \\
  ({\rm NW})&:\ (\rho,u,v,p)=(1,\,0.7276,\,0,\,1), \\
  ({\rm SW})&:\ (\rho,u,v,p)=(0.8,\,0,\,0,\,1), \\
  ({\rm SE})&:\ (\rho,u,v,p)=(1,\,0,\,0.7276,\,1),
\end{aligned}
  \label{eq:case12-ic}
\end{equation}
run to $t_{\rm end}=0.25$: two contacts (NE--NW, NE--SE), one shock (NW--SW), and one rarefaction (SW--SE), whose interaction rolls the NE--NW and NE--SE contacts up into a Kelvin--Helmholtz ``mushroom'' astride the diagonal.\\
A base grid $N_0=256$ ($\delta x_0=1/256$) carries two nested levels of refinement: level~1 at $r_1=2$ ($N=512$, $\delta x_1=\delta x_0/2$), a single, edge-connected staircase domain following the diagonal shear layer, and level~2 at $r_2=4$ relative to level~0 ($N=1024$, $\delta x_2=\delta x_0/4$), nested inside level~1 over the central mushroom roll-up. Figure~
\ref{fig:val-case12-field} shows the resulting topology along with isocontours of the density field.
To isolate the effect of the refinement itself from any confound with the sensor or the adaptive time step, we compare three runs sharing \emph{identical} solver settings and differing only in grid/refinement: a uniform grid at $N=256$ (the base resolution, 305 steps), the two-level staircase (323 level-0 steps), and a uniform grid at $N=1024$ (the finest resolution used anywhere in the staircase, 1303
steps).
\begin{figure}[h!]
\centering
\includegraphics[width=\linewidth]{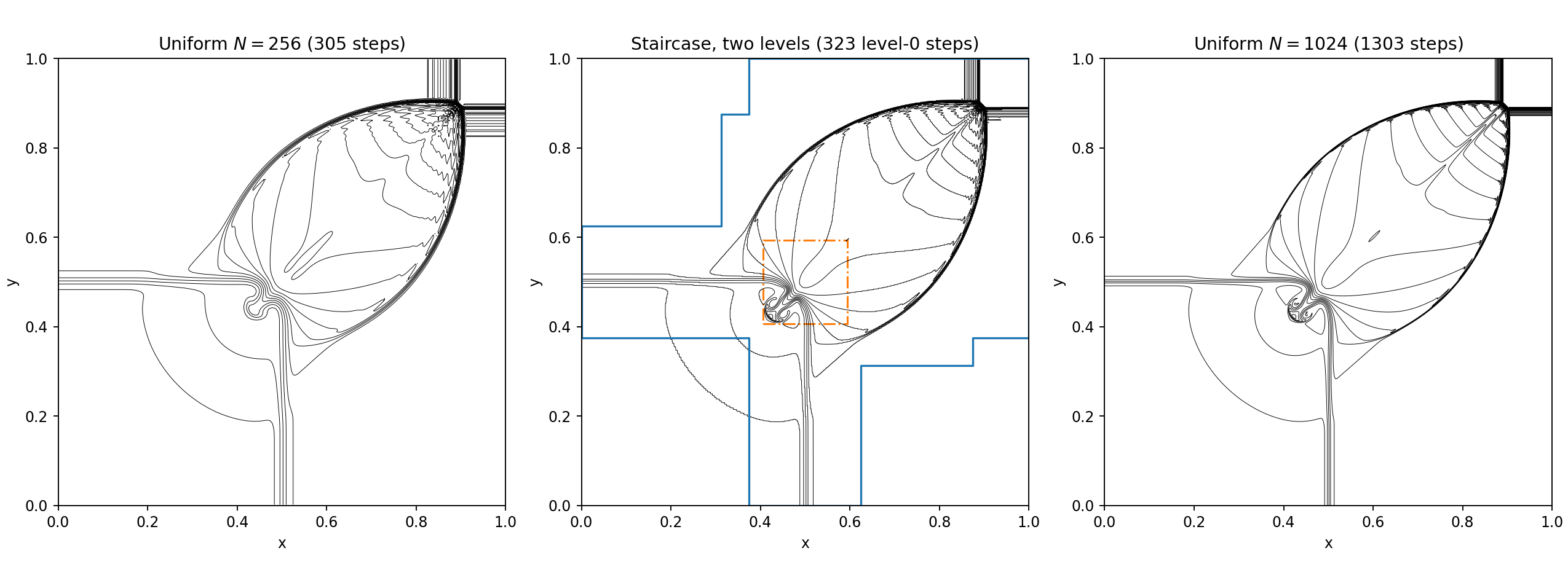}
\caption{Configuration~12, $t=0.25$, identical solver settings, varying only grid/refinement: uniform $N=256$ (left), the two-level staircase with its level-1 and level-2 footprints outlined (blue: $N=512$;
orange: $N=1024$; center), and uniform $N=1024$ (right). }
\label{fig:val-case12-isolated}
\end{figure}
Figure~\ref{fig:val-case12-isolated} shows the result: the staircase run's mushroom roll-up and diagonal shear layer are visually indistinguishable from the uniform-$N{=}1024$ reference, while the uniform-$N{=}256$ panel is markedly more diffuse in the same region, confirming that the added sharpness is attributable to the refinement itself under matched dissipation settings.
\begin{figure}[h!]
\centering
\includegraphics[width=0.6\linewidth]{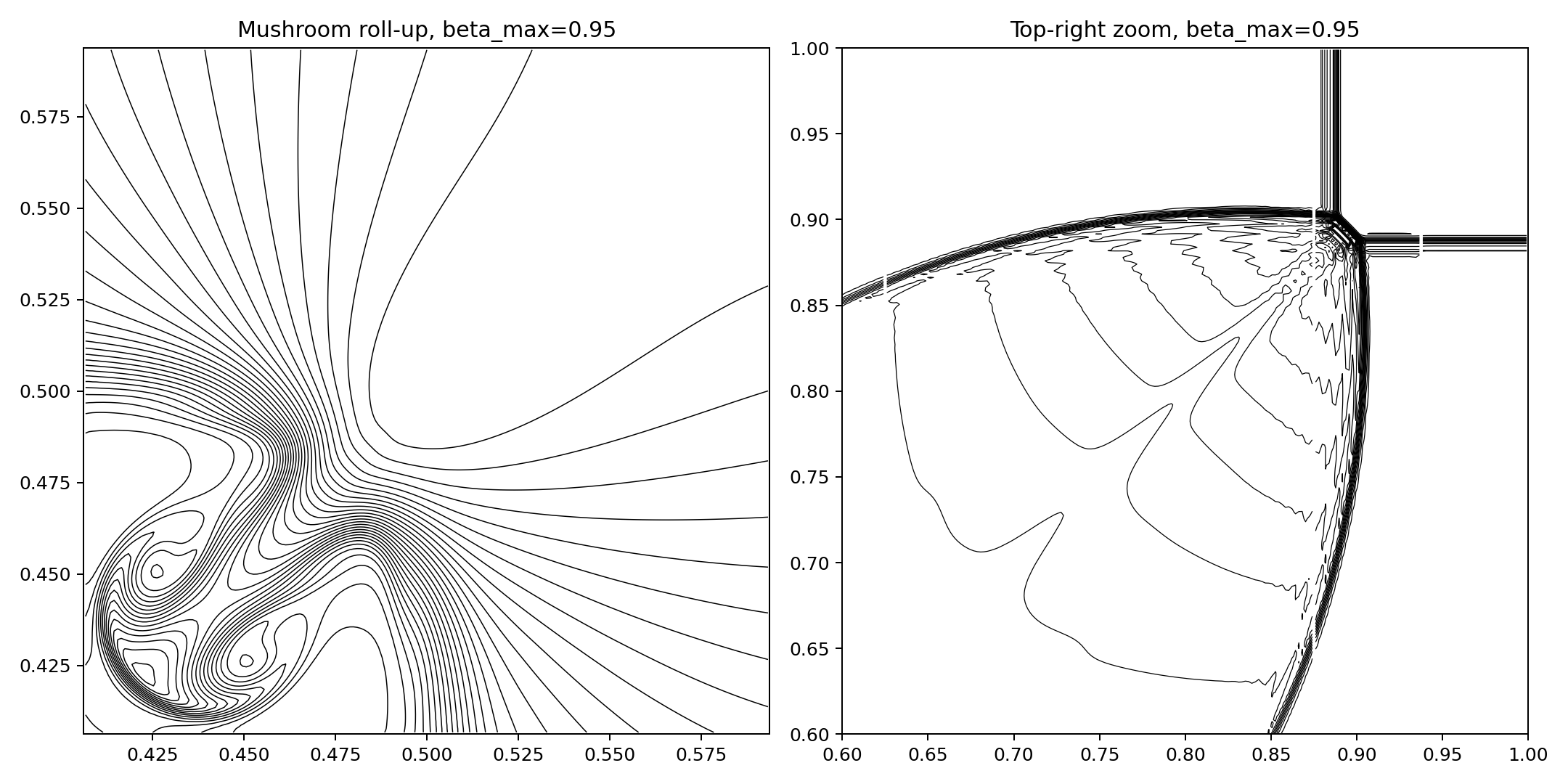}
\caption{Configuration~12, $t=0.25$: (left) the mushroom roll-up on its own $N=1024$ level-2 grid, showing closed contours in the roll-up core; (right) a zoom on the opposite corner of the staircase.}
\label{fig:val-case12-mushroom}
\end{figure}
Figure~\ref{fig:val-case12-mushroom} zooms on the mushroom core, resolved on its own $N=1024$ level-2 grid with closed contours in the roll-up. Qualitatively, the morphology of the diagonal shear layer and
its mushroom roll-up produced here closely follows that captured by reference solutions.
\subsection{Dynamic adaptive refinement: double Mach reflection}
\label{sec:val-dmr-amr}
\begin{figure}[h!]
\centering
\includegraphics[width=\linewidth]{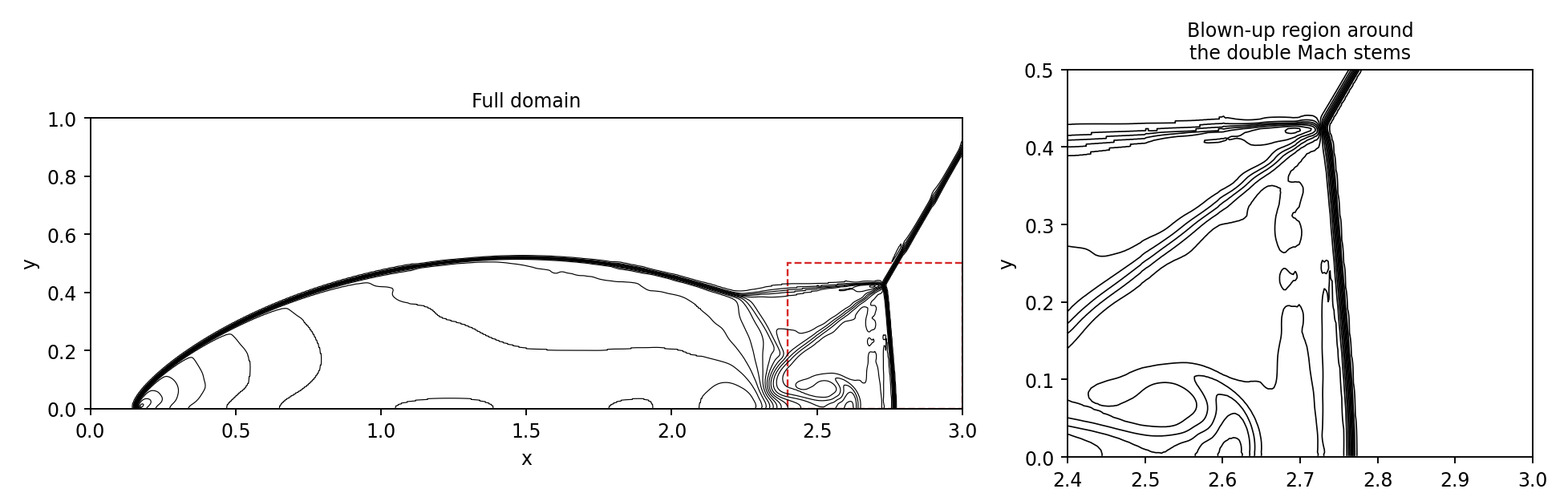}
\caption{Double Mach reflection, $t=0.200$: our result redrawn in the contour convention of Shi, Zhang and Shu~\cite{ShiZhangShu2003} ($30$ equally spaced lines, $\rho\in[1.5,22.9705]$) -- full domain (left)
with the blown-up inset's crop marked by the dashed red box, and the blown-up inset around the double Mach stems itself (right).}
\label{fig:val-dmr-wenocompare}
\end{figure}

Every case so far refines an area that is fixed for the whole run, chosen once from the geometry of the problem. The last case puts \S\ref{sec:amr} to work as intended: an adaptive refinement zone, on a genuinely unsteady two-dimensional problem whose feature of interest is a front crossing the domain at an angle. We use the double Mach reflection benchmark of Woodward and Colella~\cite{WoodwardColella1984}: a planar Mach-10 shock in a $\gamma=1.4$ gas, inclined at $60^\circ$ to the wall, initially touching $y=0$ at $x=1/6$ on a domain $[0,4]\times[0,1]$, with the exact post-shock state imposed as a time-dependent inflow along the top boundary that tracks the shock's own motion, run to $t_{\rm end}=0.2$ as is standard for this benchmark.\\
A base grid $N_0=800\times200$ ($\delta x_0=1/200$) carries two dynamically adaptive levels, each at $r=2$, up to an effective $3200\times800$ resolution ($\delta x_2=\delta x_0/4$) wherever both are active. Cells are retagged and re-clustered every $n_{\rm regrid}=5$ coarse steps from the max-norm sensor $\hat s$ of Eq.~\eqref{eq:amr-sigma} (tolerance $0.01$, dilated by a $10$-cell buffer, tiled on a $20\times12$-cell grid).
\begin{figure}[h!]
\centering
\includegraphics[width=0.7\linewidth]{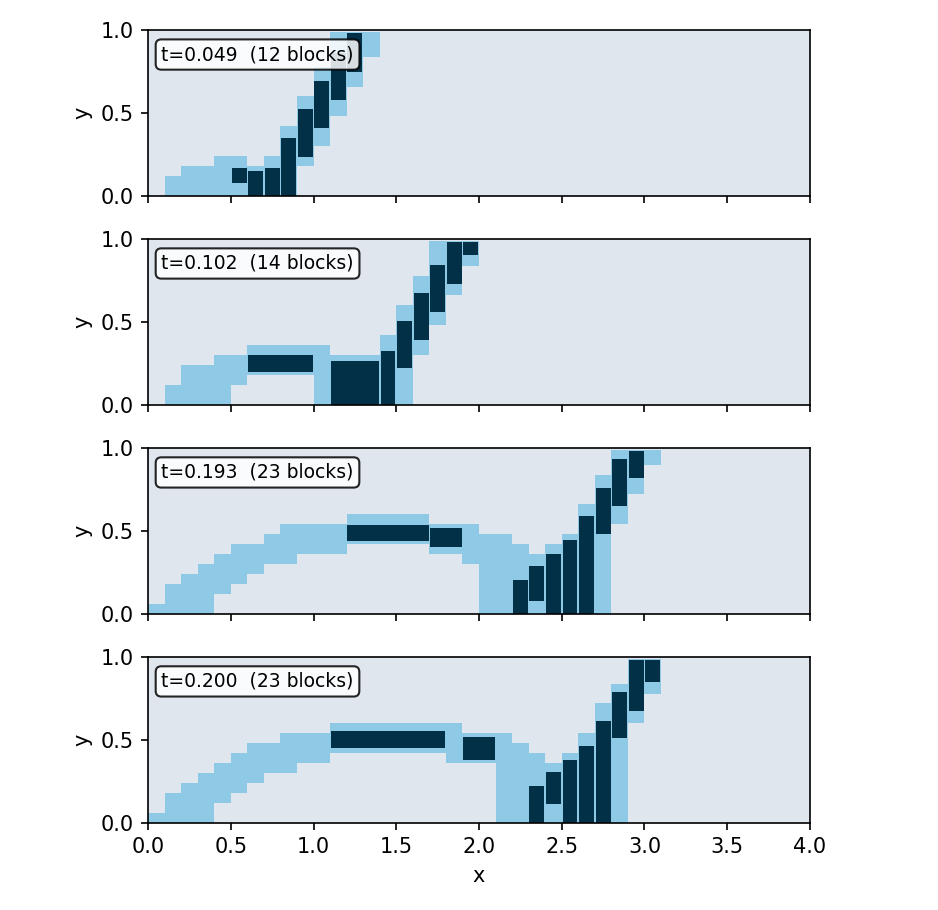}
\label{fig:val-dmr-footprint}
\caption{Double Mach reflection: the active-level map at four times during the run.}
\end{figure}
Figure~\ref{fig:val-dmr-footprint} shows the resulting refinement zones at four times.
\begin{figure}[t]
\centering
\includegraphics[width=0.9\linewidth]{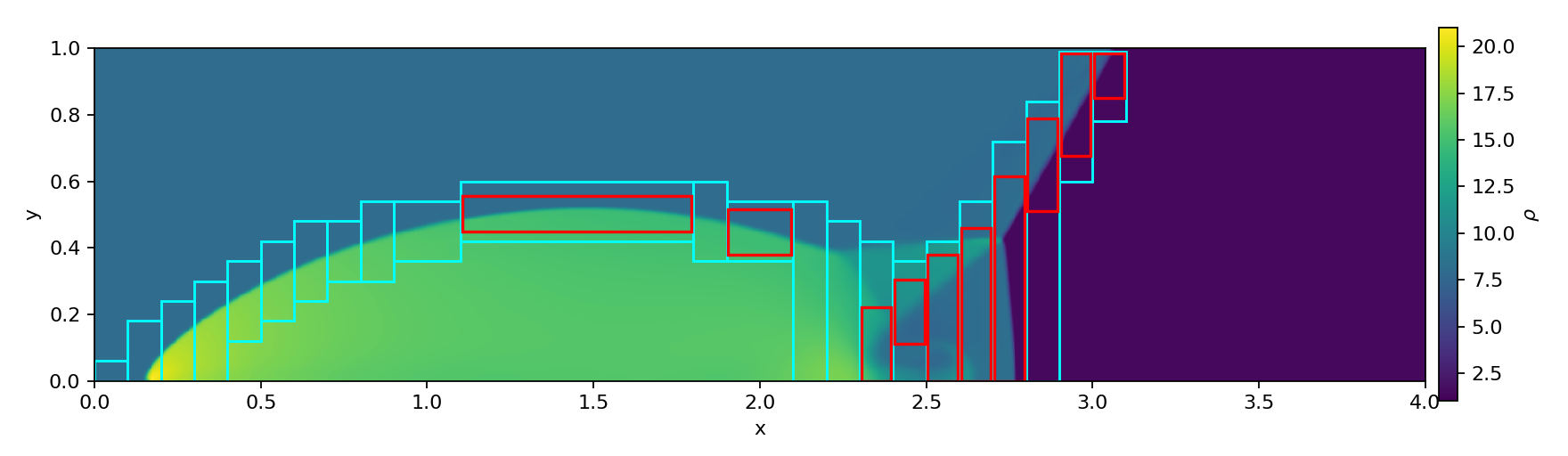}
\caption{Double Mach reflection, $t=0.200$: the $23$ level-1 blocks (cyan) outlined over the density field.}
\label{fig:val-dmr-blocks}
\end{figure}
Figure~\ref{fig:val-dmr-blocks} illustrates the times dependence of the refinement zones. A precise cell-update accounting against an equivalent uniform $3200\times800$ run is harder to state with confidence here than for the stationary-refinement cases, since the fronts being tracked occupy a genuinely large, and growing, fraction of the domain by $t_{\rm end}=0.2$; even so, the final level-1 area covers only about a quarter of the base-grid area. In addition, a comparison with fifth- and ninth-order finite-difference WENO (Weighted Essentially non-Oscillatory) on uniform meshes up to $h=1/960$ ($3840\times960$) as reported in~\cite{ShiZhangShu2003}, shows very good agreement. Fig.~\ref{fig:val-dmr-wenocompare} redraws our own result in their exact convention -- $30$ equally spaced density contour lines from $\rho=1.5$ to $\rho=22.9705$, the same domain crop and a matching blown-up inset. The number and tightness of the roll-up windings obtained here are visually consistent with the reference.

\section{Conclusion and discussion}
\label{sec:conclusion}
We have presented a conservative local grid-refinement coupling for the vectorial lattice Boltzmann method with sub-cycling in time to maintain acoustic scaling, guaranteeing convergence to the proper Euler level dynamics. A mean-preserving flux reconstruction in time supplies the coarse-to-fine boundary data, and a matched, Berger--Colella-style reflux returns the fine-to-coarse data, together conserving mass, momentum, and energy exactly, independent of the equation of state, adaptive time-stepping and relaxation coefficient. The grid refinement strategy was validated through a variety of static-refinement-based configurations all showing exact conservation and consistent convergence behavior to reference solutions, and, in \S\ref{sec:amr} and \S\ref{sec:val-dmr-amr}, extended to adaptive mesh refinement through a conservative splitting/merging pair, without any modification to the underlying coupling.

Combined with ingredients proposed in previous publications such as adaptive time-stepping and shock-capturing dissipation~\cite{StrassleHosseiniKarlin2025, Strassle2026Sensors}, the refinement strategy opens the door for more advanced multi-scale simulations with the VLBM.

\section*{Declaration of competing interest}
The authors declare that they have no known competing financial interests or personal relationships that could have appeared to influence the work reported in this paper.

\section*{Acknowledgements}
This research was funded by the Swiss National Science Foundation (SNSF) Grants 200021-228065 and 200021-236715. Computational resources at the Swiss National Supercomputing Centre (CSCS) were provided under Grants No.\ s1286, sm101 and s1327.\\
During the preparation of this work, the author(s) used Claude AI to assist with code development/data analysis and manuscript drafting. After using this tool, the author(s) reviewed and edited the content as needed and take full responsibility for the content of the published article.
\bibliographystyle{elsarticle-num}
\bibliography{aipsamp}

\end{document}